\documentclass[12pt]{article}
\newif\iffigs\figstrue
\usepackage{latexsym}
\iffigs
  \input{epsf}
\else
\fi
\font\tenmsbm=msbm10 scaled 1200
\font\sevenmsbm=msbm9
\newfam\msbmfam
\textfont\msbmfam=\tenmsbm
\scriptfont\msbmfam=\sevenmsbm
\def\msbm{\fam\msbmfam\tenmsbm}

\makeatletter
\@addtoreset{equation}{section}
\makeatother
\renewcommand{\theequation}{\thesection.\arabic{equation}}
\newcommand{\eqn}[1]{(\ref{#1})}

\newsavebox{\uuunit}
\sbox{\uuunit}
    {\setlength{\unitlength}{0.825em}
     \begin{picture}(0.6,0.7)
        \thinlines
        \put(0,0){\line(1,0){0.5}}
        \put(0.15,0){\line(0,1){0.7}}
        \put(0.35,0){\line(0,1){0.8}}
       \multiput(0.3,0.8)(-0.04,-0.02){10}{\rule{0.5pt}{0.5pt}}
     \end {picture}}
\newcommand {\unity}{\mathord{\!\usebox{\uuunit}}}

\def\IP{\relax{\rm I\kern-.18em P}}

\def\inbar{\vrule height1.5ex width.4pt depth0pt}
\def\IC{\relax\,\hbox{$\inbar\kern-.3em{\rm C}$}}
\def\bfzero{\relax\,\hbox{$\inbar\kern-.3em{\rm 0}$}}
\def\IR{\hbox{\msbm R}}
\def\ZZ{\hbox{\msbm Z}}
\def\SS{\hbox{\msbm S}}
\def\bfone{\relax{\rm 1\kern-.35em 1}}

\def\cA{{\cal A}} \def\cB{{\cal B}}

\def\cL{{\cal L}} \def\cM{{\cal M}}

\def\tilde{\widetilde}

\def\IE{\relax{{\rm I\kern-.18em E}}}

\def\IGam{\relax{{\rm I}\kern-.18em \Gamma}}

\def\bet{\begin{tabular}}
\def\eet{\end{tabular}}
\def\ol{\overline}

\def\a{\alpha}
\def\b{\beta}
\def\l{\lambda}
\def\f{\phi}

\def\g{\gamma}
\def\G{\Gamma}
\def\s{\sigma}
\def\e{\epsilon}

\def\bl{\bar{\lambda}}

\def\bpsi{\bar{\psi}}

\def\bG{\bar{\Gamma}}

\def\hs{\hat{\s}}
\def\uA{\underline{A}}
\def\ua{\underline{a}}
\def\ub{\underline{b}}
\def\uc{\underline{c}}
\def\ud{\underline{d}}
\def\ue{\underline{e}}
\def\um{\underline{m}}
\def\uaa{\underline{\a}}
\def\uab{\underline{\b}}
\def\uag{\underline{\g}}
\def\uam{\underline{\mu}}
\def\uad{\underline{\delta}}
\def\ha{\hat{a}}
\def\hb{\hat{b}}
\def\hc{\hat{c}}
\def\hd{\hat{d}}
\def\hm{\hat{m}}
\def\hn{\hat{n}}
\def\hr{\hat{r}}
\def\bpsi{\ol{\psi}}
\def\bth{\ol{\theta}_{-}}
\def\th{\theta_{-}}
\def\q{{\bf q}}
\begin{document}
\begin{titlepage}
\begin{flushright}
Preprint DFTT 98/39 \\
hep-th/9807115\\
July 1998\\
\end{flushright}
\vskip 2cm
\begin{center}
{\Large \bf  The $Osp(8|4)$ singleton action\\
\vskip 1.5mm
from the supermembrane$^*{}^\dagger$ }\\
\vfill
{  Gianguido Dall'Agata$^1$, Davide Fabbri$^1$,
   Christophe Fraser$^2$, \\
   Pietro Fr\'e $^1$, Piet Termonia$^1$
   and  Mario Trigiante$^3$   } \\
\vfill
{\small
$^1$ Dipartimento di Fisica Teorica, Universit\'a di Torino, via P. 
Giuria 1,
I-10125 Torino, \\
 Istituto Nazionale di Fisica Nucleare (INFN) - Sezione di Torino, 
Italy \\
\vspace{6pt}
$^2$ Dipartimento di Fisica Teorica, Universit\'a di Torino, via P. 
Giuria 1,
I-10125 Torino
and \\
Dipartimento di Fisica Politecnico di Torino, C.so Duca degli Abruzzi,
24,
I-10129 Torino\\
\vspace{6pt}
$^3$ Department of Physics, University of Wales Swansea, Singleton
Park,\\
 Swansea SA2 8PP, United Kingdom\\
}
\end{center}
\vfill
\begin{abstract}
Our goal is to study the supermembrane on an $AdS_4 \times {\cal 
M}_7$ background,
where ${\cal M}_7$ is a $7$--dimensional Einstein manifold with $N$ 
Killing spinors.
This is a direct way to derive the $Osp(N\vert 4)$ singleton field 
theory with all
the additional properties inherited from the geometry of the internal 
manifold.
As a first example we consider the maximally
supersymmetric $Osp(8\vert 4)$ singleton corresponding to the choice
${\cal M}_7={\msbm S}^7$. We find the explicit form of the action of 
the membrane
coupled to this background geometry and show its invariance under 
non--linearly
realized super-conformal transformations.
To do this we introduce the supergroup generalization of the solvable
Lie algebra parametrization of non--compact coset spaces.
We also derive the action of quantum fluctuations around the
classical configuration,
showing that this is precisely the singleton action.
We find that the singleton is simply realized as a free field theory
living on flat Minkowski space.
\end{abstract}
\vspace{2mm} \vfill \hrule width 3.cm
{\footnotesize
 $^*$ Supported in part by   EEC  under TMR contract
 ERBFMRX-CT96-0045, \\
 $^{\dagger}$ Supported in part by EEC  under TMR contract
 ERBFMRX-CT96-0012, in which M. Trigiante is associated to Swansea
University}
\end{titlepage}
\section{Introduction}
\label{intro}
There has recently been renewed interest in
compactified supergravity vacua of the form:
\begin{equation}
\label{adsvacua}
M_D = AdS_{p+2}\times {\cal M}_{D-p-2}
\end{equation}
where $D$ denotes the total dimension of space--time,
\begin{equation}
\label{adscoset}
AdS_{p+2} \equiv \frac{SO(2,p+1)}{SO(1,p+1)}
\end{equation}
denotes an anti de Sitter space in $p+2$ dimensions and ${\cal 
M}_{D-p-2}$
is some choice of a {\it compact Einstein manifold} in the
complementary $D-p-2$ dimensions. This interest is due to a
conjectured and partly proved {\it holographic correspondence}
between the {\it quantum dynamics}
of a {\it conformal field theory} (CFT) describing
the infrared fixed point behaviour of a gauge theory (GT)
in $p+1$ dimensions and the {\it classical tree--level
dynamics} of the {\it Kaluza--Klein supergravity theory}
(KK) obtained by compactification and harmonic expansion on
${\cal M}_{D-p-2}$. In this  correspondence, originally proposed by 
Maldacena \cite{maldapasto},
GT is the effective field theory of a large number of $Dp$--branes, 
while the $AdS_{p+2}$ metric
describes the near horizon geometry of the corresponding classical
brane solutions of D--dimensional supergravity.
This correspondence was further generalised in \cite{maldapasto}
to the worldvolume conformal field theory of $N$ coincident 
$M$--branes,
which was conjectured to be dual to $M$--theory on the 
anti de Sitter background.
\subsection{The algebraic basis of the holographic correspondence}
The algebraic basis of Maldacena's correspondence was an important 
observation recently stated in  \cite{renatoine1,renatoine2}, 
recalling the considerations made in 
\cite{BlencoweDuff,supersingleton,QFT, Nicolai} on the famous 
membrane at the end of the world.
Namely the anti de Sitter symmetry
of the bulk theory is realised as conformal symmetry on the brane
which is located at the boundary, hence {\it holography}.
 
To be explicit, consider an Mp--brane solution (where $p = 2,5$) 
of $D=11$ M--theory, namely a metric
\begin{equation}
ds^2_{11} \, = \, \left(1+
\frac{k}{r^{\tilde d}} \right)^{-\frac{\tilde d}{9}} \,
dx^I  \,   \, dx^J \eta_{IJ} + \, \, \left(1+
\frac{k}{r^{\tilde d}} \right)^{\frac{  d}{9}} \, dy^{\hat a} \,  \,
dy^{\hat b} \, \delta_{{\hat a}{\hat b}}\, .
\label{protyp}
\end{equation}
where
\begin{equation}
 d  \equiv  p+1  ; \quad  {\tilde d}  \equiv  11 - d -2
\label{ddtil}
\end{equation}
are the world--volume dimensions of the $p$--brane and of its
magnetic dual,
\begin{equation}
r \, \equiv \, \sqrt{ y^{\hat a} \,  y^{\hat b} \, \delta_{{\hat 
a}{\hat b}} }
\label{rdef}
\end{equation}
is the radial distance from the brane in transverse space,
$I,J=0,\ldots,d-1$ and ${\hat a},{\hat b}=d,\ldots,10$.
 
It has been known for some years
\cite{solitoni,doubling} that near the horizon ($ r \, \to \, 0$)
 the exact
metric \eqn{protyp}) becomes approximated by the metric of the 
following
$11$--dimensional space:
\begin{equation}
{  M}^{hor}_{p}\, =\, AdS_{p+2} \, \times \, \SS^{9-p}
\label{mh11}
\end{equation}
that has
\begin{equation}
{\cal I}^{hor}_p \,=\, SO(2,p+1) \, \times \, SO(10-p)
\label{Ihp}
\end{equation}
as  isometry group.
 
It was observed \cite{renatoine2} that the
Lie algebra of ${\cal I}^{hor}_p$ can be identified with the bosonic
sector of a superalgebra ${\cal SC}_p$ admitting the interpretation
of conformal superalgebra on the $p$--brane world--volume.
The explicit identifications are
\begin{equation}
\begin{array}{ccccccc}
{\cal I}^{hor}_2 & = & SO(2,3) \, \times \, SO(8) & , &
{\cal SC}_2 & = & Osp(8\vert 4), \\
\null & \null & \null & \null & \null & \null & \null \\
 {\cal I}^{hor}_5 & = & SO(2,6) \, \times \, SO(5) & , &
{\cal SC}_5 & = & Osp(2,6\vert 4), \\
\end{array}
\label{scgroup}
\end{equation}
where $Osp(8\vert 4)$ is the real section of the complex
orthosymplectic
algebra $Osp^c(8\vert 4)$ having $SO(8) \, \times \, Sp(4,\hbox{\msbm 
R})$ as
bosonic
sub-algebra, while $Osp(2,6\vert 4)$ is the real section of the same
complex superalgebra having $SO(2,6) \, \times \, \left(USp(4)\sim
SO(5) \right)$ as bosonic sub-algebra.
 
In \cite{renatoine2} it was shown how to realize the transformations
of ${\cal SC}_p$ as symmetries of the linearized $p$--brane
world--volume action. 
In \cite{maldapasto} it was instead suggested that
the non--linear Born Infeld effective action of the $p$--brane 
in the \eqn{mh11} background, is
invariant under conformal like transformations that realize the
group
${\cal I}^{hor}_p$. Since these transformations are similar but not
identical to the standard conformal transformations, they have been 
named
{\it broken conformal transformations}.
Further developments in this direction appeared
in \cite{holow,nairdaemi}.
 
These cases correspond to the 
choice
${\cal M}_{11-p-2} =\SS^{9-p}$ which, in the context of KK, yields 
the maximal number of preserved supersymmetries
\footnote{In the case of the $\SS^7$--compactification, the near 
horizon
bulk theory is gauged $N=8$ supergravity \cite{dewit1}}.
However other choices of ${\cal M}_{11-p-2}$ are available.
In \cite{GsuH} it was shown that there is a one--to--one
correspondence between Freund--Rubin compactifications of $D=11$
supergravity \cite{freundrub,KKduff,KKenglert,mpqr,KKwarncastel, 
dewit2} and 
Mp--brane solutions,
in the sense that the
Freund--Rubin solution on the manifold \eqn{adsvacua} with $p=2$ or
$p=5$ is the near--horizon geometry of a suitable Mp--brane for each
choice of ${\cal M}_{D-p-2}$. As KK vacuum the Freund--Rubin
solution preserves $N_{{\cal M}}$ supersymmetries
in $AdS_{p+2}$ space where, by definition,  $N_{{\cal M}}$ is
the number of  Killing spinors $\eta_A$, defined as follows:
\begin{equation}
\left [ {\cal D}^{\cal M}_m + e \Gamma _m \right ] \eta_A =0 \quad ;
\quad A=1,\dots, N_{\cal M}.
\quad
\label{kilgspi}
\end{equation}
In \eqn{kilgspi} ${\cal D}^{\cal M}_m$ is the spinorial covariant 
derivative 
on ${\cal M}$, $ \Gamma _m$ denotes the Dirac matrices in dimension
$ \mbox{dim}{\cal M}$ and $e$ is related to the $AdS$ radius. 
Hence for $p=2$ the number of supercharges preserved near the
horizon is $4 N_{\cal G}$, while in the bulk it is $1/2$
of that number, namely $2 N_{\cal G}$.
 
\subsection{$G/H$--branes}
Of particular interest are the $G/H$ branes introduced in \cite{GsuH}
and already considered in \cite{duffetal}. They correspond to the
choice of a homogeneous coset manifold as internal space:
\begin{equation}
\label{GsuHperM}
{\cal M}_{D-p-2} = \frac{G}{H} \quad ; \quad \mbox{dim} G - 
\mbox{dim}H = D-p-2
\end{equation}
and are in one--to--one correspondence with the $G/H$ Freund--Rubin
compactifications of D=11 supergravity completely classified in
\cite{KKwarncastel} and thoroughly studied in the eighties
\cite{KKduff,KKenglert,mpqr,castdauriafre}. The $7$--dimensional
coset manifolds for the $p=2$ case and the $4$--dimensional coset
manifolds for the $p=5$ case constitute a finite set and all the
$N_{G/H}$ numbers are known (see \cite{GsuH} for a summary). 
 
The case of the round and squashed seven spheres are the best known 
($N_{G/H}=8$ and
$N_{G/H}=1$) but in the eighties the Kaluza--Klein spectra have been 
systematically
derived  also for all the other solutions using the technique of 
harmonic expansions
\cite{mpqr1,mpqr2}. The organization of these spectra in 
supermultiplets
is known not only for the round $\SS^7$ \cite{gunaydin} but also for
the case of supersymmetric $M^{pqr}$ spaces
$$
 M^{pqr} \equiv  \frac{SU(3) \,\times \, SU(2) \, \times \, 
U(1)}{SU(2) \, \times \,
 U(1) \, \times \, U(1)}
$$
where $p,q,r\in \hbox{\msbm Z}$ define the embedding of the
$U(1)^2$ factor of $H$ in $G$.  For $p=q=\mbox{odd}$ we have 
$N_{G/H}=2$, in all the other (non supersymmetric cases)
we have $N_{G/H}=0$.
The $N=2$ multiplet structure was obtained
in \cite{mpqr4}.
 
Since much is known about these spaces, $G/H$--branes
constitute an excellent laboratory where to make direct checks of the
{\it holographic correspondence}:
\begin{equation}
\label{holography}
\mbox{CFT on $\partial (AdS_{p+2})$} \quad \leftrightarrow \quad
\mbox{KK on $AdS_{p+2}$}
\end{equation}
\subsubsection{The qualitative difference between the round $\SS^7$
case and the lower supersymmetry cosets $G/H$}
Let us now stress the qualitative difference between the
case with maximal supersymmetry and the cases with lower
supersymmetry. Recalling results that were
obtained in the early eighties \cite{mpqr1,mpqr2}, we know that, if
the Freund Rubin coset manifold  admits $N_{G/H}$ Killing
spinors, then the structure of the isometry group $G$ is necessarily
factorized in the following way:
$$
G \, = \, G^\prime \, \otimes \, SO\left( N_{G/H} \right)
$$
where  the $R$--symmetry factor $SO\left( N_{G/H} \right)$
can be combined with the isometry group $SO(2,3)$ of anti de Sitter
space to produce the orthosymplectic algebra $Osp \left( N_{G/H} 
\vert 4
\right )$, while the factor $G^\prime$ is the gauge--group of the
vector multiplets. Correspondingly the three--dimensional 
world--volume action
of the $CFT$ must have the following superconformal symmetry:
\begin{equation}
\begin{array}{ccc}
{\cal SC}_2^{G/H} & = & Osp \left( N_{G/H} \vert 4 \right) \, \times
\,
G^\prime \\
\end{array}
\label{scGH}
\end{equation}
where $G^\prime$ is a {\it flavour group}.
Here comes the essential qualitative difference between the maximal
and lower supersymmetry cases. In the maximal case the harmonics on 
$G/H$
are labeled only by $R$--symmetry representations while
in the lower susy case they  depend both on
$R$ labels and on representations of the gauge/flavour group 
$G^\prime$.
The structure of $Osp(8\vert 4)$ supermultiplets
determines completely their $R$--symmetry representation content so 
that the harmonic
analysis becomes superfluous in this case. The eigenvalues of the 
internal laplacians
which determine {\it the Kaluza--Klein masses} of the $Osp(8\vert 4)$ 
graviton
multiplets or, in the conformal reinterpretation of the theory, the 
{\it conformal
weights} of the corresponding primary operators, are already fixed by 
supersymmetry
and need not be calculated. In this sense the
correspondence \eqn{holography} is somewhat trivial in the maximal
susy case: once the superconformal algebra ${\cal SC}_2$ has been
identified with the super-isometry group $Osp(8\vert 4)$ the
correspondence between conformal weights and Kaluza--Klein masses is
simply guaranteed by representation theory of the superalgebra.
On the other hand in the lower susy case the structure of  the
$Osp(N_{G/H}\vert 4)$ supermultiplets fixes only their content in
$SO \left( N_{G/H} \right)$ representations while the Kaluza--Klein
masses, calculated through harmonic analysis depend also on
$G^\prime$ labels. In this case the holographic correspondence yields
a definite prediction on the conformal weights that, as far as 
superconformal symmetry
is concerned would be arbitrary. Explicit verification of these
predictions would provide a much more stringent proof of the
holographic correspondence and yield a deeper insight in its inner
working. However in order to set up such a direct verification one
has to solve a problem that was left open in Kaluza--Klein
supergravity: the singleton problem.
\subsection{The singleton problem in $G/H$ M2 branes}
As it is well known  both from 
\cite{Dirac,fronsdal1,fronsdal2,fronsdal3} and from the study of 
Kaluza--Klein supergravity in the eighties (for a review see
\cite{castdauriafre}), {\it apart from one exception}, all
the unitary irreducible representations of the  $N$--extended
anti de Sitter superalgebra correspond to supermultiplets of ordinary
fields characterized by a mass and a spin and living in the bulk of
anti de Sitter space. 
The massless representations are in one--to--one correspondence 
with the analogue massless multiplets of the
$N$--extended Poincar\'e superalgebra and in addition there is a 
wealth of shortened
massive multiplets that realize BPS saturated states of string
theory or M--theory. These ordinary short and long multiplets
appear in the Kaluza--Klein expansion of $D=11$ supergravity around
an anti de Sitter background \eqn{adsvacua}. The exception is the
lowest lying unitary irreducible representation of
the $AdS$ superalgebra, the {\sl singleton}, which does not admit a
field theory realization in the bulk of $AdS$--space but which is the
building block for all the other representations, in the sense that
all supermultiplets can be obtained decomposing tensor products of
the singleton supermultiplet.
 
In the eighties, the field theory realization of the singleton was
considered by several authors \cite{supersingleton,QFT,Nicolai,masse}.
Following previous results on the non supersymmetric
case \cite{fronsdal1,fronsdal3}, it was realized that the
singleton field theory lives on the boundary of $AdS$ space.
It was also realized that the (super) anti de Sitter
group acts as the (super) conformal
group on its boundary, and thus on the singleton field theory.
This fact led to the identification of the singleton theory with 
the world--volume conformal field theory on a brane placed at the
boundary.
 
Recently, a deeper understanding of the singleton has been promoted by
the {\it holographic correspondence} \eqn{holography}. The singleton
field theory of $AdS_{p+2}$ lives in one lower dimension 
(i.e. $d = p+1$) since it is identified with the microscopic gauge
field theory on the brane world--volume.
The tensor product
realization of the ordinary $AdS$--supermultiplets corresponds to the
construction of composite operators in the world--volume theory 
playing the 
role of emission
vertices for all KK states.
 
In the case of maximal supersymmetry there is little to discover from
the group--theoretical view--point since, as already emphasized, the
$Osp(8\vert 4)$ singleton contains  uniquely fixed
representations of the $SO(8)$ $R$--symmetry group.
Instead, in the lower susy case of $G/H$--branes there is a crucial
group--theoretical information that needs to be extracted from 
dynamics.
This  is the representations of the flavour group $G^\prime$ to which 
the
singleton has to be assigned.
Such an information, which is the prerequisite for
any verification of the {\it holographic correspondence} 
\eqn{holography}, cannot be extracted from Kaluza--Klein
supergravity but can be provided only by the world--volume field 
theory.
 
It is for this reason that in the present paper we consider the
derivation of the singleton field theory from the supermembrane
action that couples consistently to any $D=11$ supergravity
background.
 
To avoid any confusion, we would like to stress here that we call 
singleton field theory the flat space limit of the free field theory  
of \cite{QFT,Nicolai}.
We point out that, since we are going to find a theory living on a 
three--dimensional Minkowski space rather than on $\SS^2 \times \SS^1$,
we have no scalar mass term which 
was instead required in \cite{QFT,Nicolai} for conformal invariance.
We will see that it can also be derived as the theory living
on the solitonic brane of \eqn{protyp}. 
However, we will also present the derivation of the interacting 
conformal
field theory describing the dynamics of a single probe brane in the 
background
of $N$ other coincident branes ($N$ large). 
 
In \cite{Ferrara}, another
definition of the singleton field theory is given as the interacting
non-abelian conformal field theory living on the boundary, and it
is this theory which is dual to the bulk supergravity.
Nonetheless, holography is also fundamental to the derivation of the
free singleton field theory (hereafter simply referred to as the 
singleton
field theory), as we will see later.
 
Furthermore, it is worth noting that though the free singleton field 
theory 
and the non-abelian conformal field theory are apparently unrelated, 
the former does provide information about the latter in the case of 
the $G/H$ 
branes, as it tells under which representation of $G'$ 
\eqn{scGH} the singletons
transform (independently of whether they are free or interacting).
\subsubsection{The singleton from the supermembrane}
The route we follow is a priori conceptually simple. We consider the
supermembrane action invariant with respect to 
$\kappa$--supersymmetry. It
can be written in any background of the {\it elfbein} $E^{\ua}$,  of 
the
{\it gravitino } 1--form $\Psi^{\uaa}$ and of the $3$--form $A$ that 
are solutions of
$D=11$ supergravity. Specializing the background to be
$$
AdS_4 \, \times \, \left(\frac{G}{H}\right)_7
$$
we should obtain an interacting conformal field theory. Indeed after 
fixing reparametrization invariance which removes $3$ of the $11$ 
bosonic
coordinates and after gauge fixing $\kappa$ supersymmetry which
removes $16$ of the $32$ degrees of freedom we are left with
$8$ bosons and $8$ fermions (on shell) which is the field content of 
the
singleton field theory. The action must then be expanded around 
a classical solution, preserving the $AdS$ (i.e. superconformal)
symmetry. This is the free field limit, yielding the singleton theory.
 
What is far from being trivial are the
details along the route. There are three main questions one has to
address in this programme:
\begin{enumerate}
\item The identification of the boundary on which the brane lives.
\item The choice of a suitable parametrization of anti de Sitter
superspace.
\item How to expand the non--linear gauge--fixed action around a
classical solution to obtain the unitary irreducible singleton
representation.
\end{enumerate}
Although our final goal is a description of $G/H$--branes, in the
present paper we consider the solutions of the above problems in the
case of maximal supersymmetry, namely for the supermembrane on the 
following
background:
$$
AdS_4 \, \times \, \SS^7.
$$
As already emphasized, many aspects of the connection between the 
$M2$--brane and the
singleton have been studied in the past 
\cite{solitoni,supersingleton,QFT,fund}.
However, an exact derivation of the singleton action from
the supermembrane on this background has revealed
to be problematic \cite{QFT} and missed for a long time.
Therefore what we do in this paper must be viewed both as a solution 
of
so far unresolved questions and as a preparation to extract
the $G^\prime$ representation content of singleton field theories in
the case of $G/H$--branes.
\subsubsection{The  organization of the paper}
In section \ref{due}, we give our parametrization of the 
near--horizon 
geometry.
 
In section \ref{tre} we present the rheonomic construction of the
$\kappa$--supersymmetric supermembrane action in first order
formalism.
 
Next, in section \ref{quattro}, we present the construction of the 
$\kappa$-fixed
$M2$--brane action in the $AdS_4 \times \SS^7$ background comparing 
our
result with the previously found partial results.
We also elucidate the non-linear realization of the action of the
superconformal $Osp(8|4)$ group on the fields and coordinates of the
membrane inherited from the isometries of the background.
 
In section \ref{cinque}, we expand the fields in small fluctuations 
normal to the membrane, and thus we find the action of the singleton.
 
Finally, of the three appendices, {\bf A} contains our conventions,
including those of the normal coordinate expansion,
{\bf B} gives details about the conformal structure and
 the topology of the boundary, while {\bf C} contains  a detailed
explanation of the solvable parametrization of $AdS_4$ as a coset
space.
\section{The near--horizon geometry}
\label{due}
$M$--branes are classical solutions of eleven dimensional 
supergravity with the \eqn{protyp} metric.
In particular, the  $M2$--brane metric is obtained from 
\eqn{protyp} setting
$d=3$, ${\tilde d}=6$:
\begin{equation}
\label{M2metric}
ds^2 = \left( 1 + \frac{k}{r^6} \right)^{-2/3} dx^I dx^J \eta_{IJ}
+  \left( 1 + \frac{k}{r^6} \right)^{1/3} dy^{\ha} dy^{\hb} 
\delta_{\ha\hb},
\end{equation}
where now $I,J=0,1,2$ and $\ha,\hb=3,\dots,10$.
The membrane is obviously located at $r=0$.
 
It is now interesting to study the geometry near the brane.
In the  $r\to 0$ limit, the metric \eqn{M2metric} becomes
\begin{equation}
\label{orfan1}
ds^2 = \left(\frac{r}{R}\right)^4 dx^I dx^J \eta_{IJ}
+  \left(\frac{R}{r}\right)^2 dy^{\ha} dy^{\hb} \delta_{\ha\hb},
\end{equation}
where $k = R^6$.
Setting $\displaystyle \rho = \left(\frac{r}{R}\right)^2$ and using
$dy^{\ha} dy^{\hb} \delta_{\ha\hb} = dr^2 + r^2 d\Omega_7^2$,
where $d\Omega_7^2$ is the invariant metric on the sphere,
the near--horizon metric can be seen explicitly to reduce to  
$AdS_4\times \SS^7$ in horospherical $\times$ hyperspherical
coordinates \cite{renatoine2}:
\begin{equation}
\label{nearhor}
ds^2 = \rho^2 \left( -dt^2 + dx^2 + dw^2 \right) 
+  \frac{R^2}{4} \frac{1}{\rho^2} d\rho^2 + R^2 d\Omega_7^2.
\end{equation}
It must be noted that the ratio of the $AdS$ radius to the $\SS^7$ 
radius
is fixed to be $1/2$, and that after a redefinition of $\rho$ 
absorbing
an $R/2$ factor, the metric has a $R^2/4$ factor in front which can be
conformally scaled away retrieving the $AdS$ space in the solvable
parametrization \cite{GsuH}:
\begin{equation}
\label{solvmet}
d\tilde{s}^2 =  \rho^2 \left( -dt^2 + dx^2 + dw^2 \right) + 
\frac{1}{\rho^2} d\rho^2 + 4 d\Omega_7^2.
\end{equation}
 
An interesting point to note is that this approximated metric is in
fact also an exact supergravity solution \cite{freundrub} and a 
stable quantum vacuum \cite{kallosh2}.
The metric near the brane is locally that of an $AdS$ space. 
The supergravity solution fixes the local metric, but there is still a
certain arbitrariness in the choice of underlying global topology.
 
Anti de Sitter space $AdS_4$ can be represented as the hyperboloid
given by the following algebraic locus in $\IR^5$:
\begin{equation}
Y^2_0 + Y^2_1 - Y^2_2 - Y^2_3 - Y^2_4 = 1\ .
\label{quadric}
\end{equation}
We can partly parametrize this space with our coordinates 
$\rho,t,w,x$ as follows: 
\begin{equation} \label{eq:diffeo}
\left\{
\begin{array}{ccl}
Y_0 &=& \rho \, t\ , \\
Y_1 &=&\displaystyle \frac{1}{2} \left[ \rho + \frac{1}{\rho} +
\rho\left(-t^2 + w^2 + x^2\right)\right]\ ,\\
\label{solut2}
Y_2 &=& \rho \, w\ , \\
Y_3 &=& \displaystyle \frac{1}{2} \left[ \rho - \frac{1}{\rho} -
\rho\left(-t^2 + w^2 + x^2\right)\right] \ ,\\
Y_4 &=& \rho \, x\ ,
\end{array}\right.
\end{equation}
with
$$
\cases{ \rho  \, \in \ ]0,\infty [\cr
t,w,x \, \in \ ]\!-\!\infty , \infty [ }
$$
where the coordinate range is a physical choice.
Note that these coordinates do not parametrize the  whole of the
hyperboloid, but they are good coordinates for $AdS_4/\ZZ_2$
\footnote{actually they cover only a part of $AdS_4/\ZZ_2$: the one 
for
which $Y_1+Y_3\neq 0$.}, where $\ZZ_2$ acts
as the inversion $Y \to -Y$.
 
Looking back at the metric \eqn{nearhor}, sections with $\rho$ fixed 
are locally isomorphic to flat Minkowski $M_3$.
Furthermore, there exists an infinite number of classical solutions to
the $D=11$ brane-wave equations with $\rho=const,\ y^{\ha}=const$  
with
Minkowski topology \cite{renatoine2}.
The $M2$--brane of \eqn{M2metric} is just one of these membranes at 
$\rho$
fixed which has been pushed to $\rho\to 0$, that is part of the 
boundary of
our space, but we should recover a proper CFT taking the $\rho \to 
\infty$ limit.
Indeed it has to be noticed that this latter provides us with a theory on 
a conformally invariant support \cite{holow}. 
The first one instead yields a theory formulated at the horizon 
(which is not an invariant support), even if it realises the bulk 
invariances on the fields such that what we obtain is a conformally 
invariant theory.

For more details on its topology and its relation with the conformal 
boundary of $AdS$ we refer the reader to Appendix B. 
\section{The supermembrane and $\kappa$--supersymmetry}
\label{tre}
It has been known for a long time that eleven dimensional supergravity
can naturally be compactified on $AdS_4 \times \SS^7$, the simplest of
the Freund--Rubin compactifications where the four--form field 
strength
has non vanishing vacuum expectation value \cite{freundrub}. It is 
also
known that this Freund--Rubin type solution can be seen as a
consistent quantum vacuum of $D = 11$ membrane theory.
Recently \cite{kallosh2}, it has been shown that the classical 
equations of
motion of the effective theory of M--theory, namely $D=11$
supergravity, evaluated on this background, cannot receive quantum
corrections which are compatible with supersymmetry.
This means that the $AdS_4 \times \SS^7$ vacuum is described by a
fixed point where all the torsion, curvature and four form
components are covariantly constant.
$AdS_4 \times \SS^7$ is an exact solution of M--theory.
We thus propose to find an explicit expression for the membrane
world-volume theory in this background.
This action should  display
superconformal symmetry, which it inherits from the $AdS$
symmetry group of the background. This is so because the
world-volume action is a generalized $\s$--model whose target space
fields are the background fields.
Then we build a three--dimensional interacting conformal field
theory, the interactions describing the membrane dynamics.
 
To this effect we need to start from the $\kappa$ supersymmetric
action of the $D=11$ supermembrane.
Although this latter has been derived long ago \cite{supermembrane},
we devote the next subsection to such a construction, because in the
rheonomic first--order formalism the action of $\kappa$--supersymmetry
becomes particularly simple and implementing its restriction to a
specific background is very easy and clear.
This formalism is equivalent to the geometric approach, from which it 
has
been derived the action of the supermembrane on flat Minkowski space
\cite{ukra}.
We point out that this approach has been used also to derive the 
actions
for $Dp$--branes on a generic curved super background \cite{ukrb}.
\subsection{The first order ``Polyakov'' action of the supermembrane
from rheonomy }
 
The starting point for the formulation of the supermembrane action is
the geometry of superspace and the rheonomic parametrization of the
supergravity curvatures. These were obtained at the beginning of the
eighties in \cite{riccapiet}. The field content of  $D=11$
supergravity is given by the following set of exterior forms: the 
vielbein 1-form
$E^{\ua}$ ($\ua = 0,1,\dots,10$), the spin--connection  1-form
$\omega^{\ua\ub}$,  the gravitino fermionic 1--form $\Psi^{\uaa}$ 
($\uaa  = 1,\dots,32$)
and the $3$--form $A$. The first three items in the above list
constitute the dual description of the superPoincar\'e algebra in
D=11. The last item, namely the $3$--form $A$ extends it to a
free--differential algebra which could be further enlarged by the
addition of a $6$--form $\tilde{A}$ whose field strength turns out to 
be the dual of that of $A$ upon implementation of the Bianchi 
identities
\cite{pietd11}.
The definition of the $D=11$ curvatures is
\begin{eqnarray}
R^{\ua} &\equiv& dE^{\ua} - \omega^{\ua\ub} \, \wedge \, E^{\uc}  \,
\eta_{\ub\uc} - \frac{\rm i}{2} \, {\bar \Psi} \wedge \IGam^{\ua}
\Psi, \nonumber\\
\label{curvatdef}
\rho &\equiv& d\Psi - \frac{1}{4}\, \omega^{\ua\ub} \, \wedge \,
\IGam_{\ua\ub} \, \Psi, \\
R^{\ua\ub}&\equiv& d\omega^{\ua\ub} - \omega^{\ua\uc} \, \wedge \,
\omega^{\ud\ub} \, \eta_{\uc\ud}, \nonumber \\
F[A]&\equiv&dA - \frac{1}{2} \, \, {\bar \Psi} \wedge \IGam^{\ua\ub}
\Psi \, \wedge \, E_{\ua} \, \wedge \, E_{\ub}, \nonumber
\end{eqnarray}
and the corresponding rheonomic solution of the superspace Bianchi
identities is as follows:
\begin{eqnarray}
R^{\ua} &=& 0, \nonumber\\
F[A]&=&F_{\ua_1,\dots\,\ua_4} \,E^{\ua_1} \,\wedge \dots \wedge \, 
E^{\ua_4}, \label{curvatpar} \\
\rho &=& \rho_{\ua\ub} \,E^{\ua} \, \wedge \, E^{\ub}
+ \frac{\rm i}{3} \left( \IGam^{\ub_1\ub_2\ub_3} \,
F_{\ua\ub_1\ub_2\ub_3} - \frac{1}{8} \IGam_{\ua\ub_1\dots\ub_4} \,
F^{\ub_1\dots\ub_4} \, \right) \, \Psi \, \wedge \, E^{\ua}, 
\nonumber\\
R^{\ua\ub} &=& \mbox{as determined by second order formalism.} 
\nonumber
\end{eqnarray}
From these parametrizations one immediately obtains the
supersymmetry transformations as superspace Lie derivatives 
\cite{castdauriafre}:
\begin{eqnarray}
\delta \, E^{\ua} &=& \mbox{\rm i} {\bar \epsilon} \, \IGam^{\ua} \, 
\Psi, \nonumber\\
\label{ordsusy}
\delta \,\Psi &=& {\cal D} { \epsilon} \, -
 \frac{\rm i}{3} \left( \IGam^{\ub_1\ub_2\ub_3} \,
F_{\ua\ub_1\ub_2\ub_3} - \frac{1}{8} \IGam_{\ua\ub_1\dots\ub_4} \,
F^{\ub_1\dots\ub_4} \, \right) \, \epsilon\, E^{\ua}, \\
\delta \, A & = & - \mbox{\rm i} {\bar \epsilon} \, \IGam^{\ua\ub}
\Psi \, \wedge \, E_{\ua} \, \wedge \, E_{\ub} \nonumber
\end{eqnarray}
The basic idea to obtain the $\kappa$-supersymmetric action of the
supermembrane is the following. We introduce a {\it dreibein} $e^i$
($i=0,1,2$) on the three--dimensional world--volume and we write the 
following first--order action functional
\begin{equation}
\label{action}
S = \int {\Pi_{\ua}}^i E^{\ua} \wedge e^j \wedge e^k \e_{ijk} + 
\alpha_1 \,
\int {\Pi_{\ua}}^l
{\Pi_l}^{\ua} e^i \wedge e^j \wedge e^k \frac{\e_{ijk}}{3!} +
\alpha_2 \,
\int e^i  \wedge e^j \wedge e^k \frac{\e_{ijk}}{3!} +  \alpha_3 \, \q 
\int A
\end{equation}
where $\q= \pm 1$ is the ``membrane charge'', while
$\alpha _1,\alpha _2,\alpha _3$ are three real parameters to be
determined by the following two conditions:
\begin{enumerate}
\item The variational equation in the 0--form $\Pi_{\ua}^{\ i}$ must
impose its identification with the projection of the target {\it 
elfbein}
$E^{\ua}$ onto the world--volume {\it dreibein} $e^i$, namely:
\begin{equation}
E^{\ua} = e^i \Pi_i{}^{\ua}.
\label{orfan3}
\end{equation}
\item The action should be invariant against
$\kappa$--supersymmetry transformations. These are nothing else but 
ordinary
supersymmetries of the background fields, as defined by 
eq.\eqn{ordsusy},
with, however, a restricted supersymmetry parameter $\epsilon$.
The restriction corresponds to a world volume
projection that halves the $32$ components of the spinor parameter.
Explicitly this is realized by setting:
\begin{eqnarray}
\epsilon &=& \frac{1}{2} \left( 1 -\q\mbox{\rm i} {\bar \IGam} 
\right)\, \kappa, \nonumber \\
{\bar \IGam} &\equiv & \frac{\e^{ijk}}{3! \sqrt{-h}} \IGam _{ijk} =
\frac{\e^{ijk}}{3! \sqrt{-h}} \Pi_i{}^{\ua}  \Pi_j{}^{\ub}
\Pi_k{}^{\uc} \IGam_{\ua\ub\uc},
\label{projector}
\end{eqnarray}
where $\kappa$ is a free $32$--component spinor while
the symbols $h_{IJ}$, $h$ denote  the
world--volume metric and its determinant, respectively.
\end{enumerate}
The invariance with respect to $\kappa$ supersymmetry can be
realized if $\alpha_2,\alpha_3$ have suitable values and if a 
suitable $\kappa$--variation of the world--volume vielbein is 
introduced. 
Before entering further
details it is worth discussing how the $\kappa$--variation has to be
conceived with respect to the world--volume fields.
The action \eqn{action} defines a $\sigma$--model and
the field configurations are embeddings
$$
{\cal WV}_3 \, \hookrightarrow \, {\cal SP}_{11\vert 32}
$$
of the bosonic three--dimensional world--volume ${\cal WV}_3$ into the
$11\oplus 32$--dimensional superspace ${\cal SP}_{11\vert 32}$.
Hence, given an explicit coordinatization of ${\cal SP}_{11\vert 
32}$,  both the $11$
bosonic coordinates $X^{\ua} $ and the $32$ fermionic coordinates
$\Theta^{\uaa}$ become  fields depending on $\xi^0,\xi^1,\xi^2$, the
three world--volume coordinates. The ordinary supersymmetry variations
\eqn{ordsusy} are given by Lie derivatives and correspond to
fermionic diffeomorphisms in superspace:
\begin{eqnarray}
\delta X^{\ua}&=& \epsilon^{\uaa} \, k^{\ua}_{\uaa} (X,\Theta)
\nonumber \\
\delta \Theta^{\uaa}&=& \epsilon^{\uab} \, k^{\uaa }_{\uab} (X,\Theta)
\label{orfan5}
\end{eqnarray}
where $ k^{\ua}_{\uaa} (X,\Theta) ,  k^{\uaa }_{\uab} (X,\Theta) $
are, in general, functions of all the coordinates $X,\Theta$. Their
explicit form depends on the choice of the coordinate frame. Replacing
$\epsilon$ with its projected counterpart \eqn{projector} what we
said of ordinary supersymmetry holds true also for
$\kappa$--supersymmetry. Hence the important point to be
stressed is that the explicit form of $\kappa$--supersymmetries
depends on the choice of the coordinate frame for superspace and can 
be either more or less involved.  However,
adopting the rheonomic point of view, the invariance of the action 
can be established in
a way completely independent from such an explicit form.
 
The curvature definitions \eqn{curvatdef} and their rheonomic
parametrizations \eqn{curvatpar} are based on a ``mostly minus'' flat 
metric $\eta_{\ua\ub} = \mbox{diag}(+,-,-,\dots,-)$ and on gamma matrices
$\IGam^{\ua}$ generating the Clifford algebra $ \left\{ \IGam^{\ua} \, ,
\, \IGam^{\ub} \right \} \, = \,2\, \eta^{\ua \ub} $.
In the context of $p$--brane solutions it is more convenient to use a
``mostly plus'' flat metric $\eta^\prime_{\ua\ub} = 
\mbox{diag}(-,+,+,\dots,+)$.
A convenient formulation of eleven--dimensional supergravity using this
metric is the superspace formulation introduced in \cite{vincoli}
\footnote{For a detailed discussion of the notations and conventions 
we refer the reader to  Appendix A.}, where the torsions and curvatures
are defined as:
\begin{equation}
\label{defo}
\begin{array}{rcl}
T^{\ua} &\equiv& d E^{\ua} + \omega^{\ua}_{\ \ub}E^{\ub}\,,\\
T^{\uaa} &\equiv&\displaystyle d E^{\uaa} + \frac{1}{4}
\omega^{\ua \ub}(\Gamma_{\ua \ub})^{\uaa}_{\ \uab}E^{\uab}\,,\\
H &\equiv& dB\,,\\
R^{\ua \ub} &\equiv& d \omega^{\ua\ub}+\omega^{\ua}_{\ \uc}\omega^{\uc\ub}\,,
\end{array}
\end{equation}
on which one imposes the set of constraints
\begin{eqnarray}
T^{\ua} &=& E^{\uaa} E^{\uab} \Gamma^{\ua}_{\uaa\uab}; \nonumber \\
T^{\uaa} &=& \frac{1}{2} E^{\ua} E^{\ub} T_{\ub\ua}{}^{\uaa} +
E^{\uab} E^{\ua} \left( 8 \Gamma^{[\underline{3}]}{}_{\uab}{}^{\uaa}
H_{\ua [\underline{3}]} + \Gamma_{\ua 
[\underline{4}]}{}_{\uab}{}^{\uaa}
 H^{[\underline{4}]}\right) ; \nonumber \\
\label{3}
H &=& \frac{1}{4!} E^{\uc_1} \ldots E^{\uc_4} H_{\uc_4 \ldots \uc_1}
- \frac{1}{4!4!} E^{\ub} E^{\ua} E^{\uab} E^{\uaa}
\Gamma_{\ua\ub\;\uaa\uab}; \\
R_{\uc\ud} &=& \frac{1}{2} E^{\ua} E^{\ub} R_{\ub\ua,\uc\ud} +
2 E^{\uaa} E^{\uab}  \left( 4! \Gamma^{\ua\ub}_{\uab\uaa}
H_{\ua\ub\uc\ud} + \Gamma_{\uc\ud [\underline{4}]}
H^{[\underline{4}]}\right) +  \nonumber\\
&+& E^{\ua} E^{\uaa} \left(2 {T_{\ua [ \uc}}^{\uag}
\Gamma_{\ud]\uag\uaa} - {T_{\uc\ud}}^{\uag} \Gamma_{\ua\;\uag\uaa} 
\right). \nonumber
\end{eqnarray}
Thus, now the $\kappa$--symmetry variations of the fields are
\begin{eqnarray}
\delta_{\kappa} E^{\ua} &=&  2
E^{\uaa} \Gamma^{\ua}_{\uaa \uab} \left [\frac{1}{2}\,(1 + \q\bG) 
\, \kappa\right ]^{\uab}, \nonumber \\
\label{orfan6}
\delta_{\kappa} B &=& - \frac{2}{4!4!} E^{\ub} E^{\ua} E^{\uaa}
\Gamma_{\ua\ub\;\uaa\uab} \, \left[\frac{1}{2}\,(1 + \q\bG) \kappa
\right]^{\uab} \\
\delta_\kappa \, e^i & = & \frac{1}{(3-t)(t+1)} \left( 2 h^{ij} + 
(t-1) 
\eta^{ik} \right) \left\{ 2 \left[\frac{1}{2}\,(1 + \q\bG) \kappa
\right]^{\uaa}\G_{k\,\uaa\uab} E^{\uab} 
\left( \delta_j{}^k - \sqrt{-h} h^{kl} \eta_{lj} \right)\right. 
\nonumber \\
&-& \left. \left[\frac{1}{2}\,(1 + \q\bG) \kappa
\right]^{\uaa} \G_{kl\,\uaa\uab} E^{\uab} \e^{rkl} \left( 
\eta_{rj} + 
\frac{h_{rj}}{\sqrt{-h}}\right)\right\}, \nonumber
\end{eqnarray}
where the operator
\begin{equation}
\label{bG}
\bG \equiv \frac{\e^{ijk}}{3! \sqrt{-h}} \Gamma_{ijk} =
\frac{\e^{ijk}}{3! \sqrt{-h}} \Pi_i{}^{\ua}  \Pi_j{}^{\ub}
\Pi_k{}^{\uc} \Gamma_{\ua\ub\uc}
\end{equation}
has the property $\bG^2 = \unity$, 
$$
h_{ij}\equiv{\Pi_i}^{\ua}{\Pi_j}^{\ub}\,\eta_{\ua\ub}
$$
is the off-shell brane metric in flat indices and $t$ its trace.
It can be seen that the action \eqn{action} is invariant under
$\kappa$--symmetry transformations if 
\begin{equation}
\alpha _1=-1 \quad ; \quad \alpha _2=-1 \quad ; \quad \alpha _3 = -4!
\label{alpvalues}
\end{equation}

\medskip

It can be understood that the language of superspace constraints is
completely isomorphic to the rheonomic formalism.
Indeed superspace constraints and rheonomic parametrizations are just
different names for the same equations.

If one introduces the relations
\begin{equation}
\begin{array}{ccccccc}
&&\Gamma^{\ua} &=& \mbox{\rm i} \, \IGam^{\ua}&&\\
E^{\uaa} & \equiv & \displaystyle \frac{1}{\sqrt{2}} \Psi^{\uaa}  & ; 
&
B \equiv  &  =- \displaystyle \frac{1}{4 !} \, A , & H \ = \ dB,   \\
T^{\uaa} & \equiv & \displaystyle \frac{1}{\sqrt{2}} 
\, \rho^{\uaa}  & ; & H_{\ua_1\dots\ua_4} &=& 
\displaystyle - \frac{1}{4 !} \, F_{\ua_1\dots\ua_4},  
\end{array}
\label{newnames}
\end{equation}
it can be checked that the rheonomic parametrizations 
\eqn{curvatpar} and
the curvature definitions \eqn{curvatdef} exactly translate into
\eqn{defo} and \eqn{3}.
Furthermore \eqn{orfan6} can be similarly translated into standard rheonomic 
formulae for the $\kappa$--symmetry variation of all the fields and the first
two equations exactly reproduce the supersymmetry variations \eqn{ordsusy} of 
the background fields; the variation of the world--volume dreibein $e^i$
in \eqn{orfan6} is the only novelty.
\subsection{The bosonic equations of motion}
\label{boseeqn}

A useful exercise to derive the existence of static membranes and later
for the linearisation around such configurations is to vary the first
order action \eqn{action} to obtain the equations of motion of the bosonic 
fields.

Set then the fermionic coordinates to zero, we parametrize the metric 
\eqn{nearhor} with the following vielbeins
\begin{equation}
\label{P1}
\begin{array}{ccc}
E^i &=& \rho d\xi^I \delta_I{}^i \\
E^{\bullet} &=&\displaystyle \frac{R}{2} \frac{d \rho}{\rho}\\
E^{\ha} &=& {\cB}^{\ha}_{\hm}(y) dy^{\hm}
\end{array}
\end{equation}
where $\cB^{\ha}_{\hm}$ is a proper choice of vielbeins for the internal 
manifold $G/H$, and
\begin{equation}
\label{P2}
B = E^i \wedge E^j \wedge E^k \frac{\e_{kji}}{3!}.
\end{equation}
As already said, the variation with respect to $\Pi_i{}^{\ua}$ yields the
embedding equation
$$
E^{\ua} = e^i  \Pi_i{}^{\ua}.
$$

Chosen to fix the world--volume frame in terms of the target space as 
\begin{equation}
\label{frame}
e^i = E^i, 
\end{equation}
varying the action w.r.t. $e^i$ one gets the ``stress--energy tensor'' 
constraint:
\begin{equation}
\label{stress}
g_{\hm\hn} (y) \, \partial_I y^{\hm} \partial_J y^{\hn} = \frac{R^2}{4} 
\frac{1}{\rho^2} \, \partial_I \rho \partial_J \rho,
\end{equation}
which is just the analogue in supermembrane theory of the Virasoro constraint
of string theory.

This fixes the form of $\Pi_{\ua}{}^i$ in terms of $\rho$, $y^{\hm}$ and $x^I$:
\begin{equation}
\label{B}
\begin{array}{rcl}
\Pi_{j}{}^i &=& \delta_j{}^i \\
\Pi_{\bullet}{}^i &=&\displaystyle \frac{R}{2} \frac{\partial_I 
\rho}{\rho^2} \delta_I{}^i \\
\Pi_{\ha}{}^i &=&\displaystyle {\cB}_{\hm\, \ha} \frac{1}{\rho} \,
\partial_I y^{\hm} 
\delta_I{}^i
\end{array}
\end{equation}

All these equations are then useful to obtain the equations of 
motion of the bosonic fluctuations of the brane $\rho$ and $y^{\hm}$.
The variation w.r.t. $\rho$ yields:
\begin{eqnarray}
\delta S &=& \int \left\{ -\frac{R}{2} \Pi_{\bullet}^{\ i}
\frac{\partial_I \rho}{\rho^2}
d\xi^I\wedge e^j\wedge e^k \e_{ijk} \right\}
\delta \rho - \int d \left\{ \frac{R}{2} \Pi_{\bullet}^{\ i}
\frac{1}{\rho} e^j\wedge e^k \e_{ijk} \right\} \delta \rho\nonumber\\
\label{varrho}
&+& \int \left\{ \Pi_l^{\ i} \, \delta_I{}^l d\xi^I\wedge e^j\wedge e^k
\e_{ijk} \right\} \delta \rho
-\q\int \left\{ \rho^2 d\xi^I \wedge d\xi^J \wedge d\xi^K
\e_{IJK} \right\} \delta \rho\ = 0
\end{eqnarray}
and thus, substituting \eqn{frame} and \eqn{B} into \eqn{varrho} 
we get the following non-linear equation for $\rho$
\begin{equation}
\Box \rho - \frac{12}{R^2} (1 - \q) \rho^3 =0\,,
\end{equation}
where $\Box \equiv \eta^{IJ}\partial_I\partial_{J}$.

In the same way, from the variation with respect to $y^{\hm}$ we obtain:
\begin{equation}
\label{vary}
\delta S = \int \left\{ \Pi_{\ha}^{\ i}
\frac{\partial {\cal B}_{\hn}^{\ \ha}}{\partial y^{\hm}}
dy^{\hn} \wedge e^j \wedge e^k \epsilon_{ijk}\right\} \delta y^{\hm}
-\int d\left\{ \Pi_{\ha}^{\ i}{\cal B}_{\hm}^{\ \ha}
e^j \wedge e^k \epsilon_{ijk} \right\} \delta y^{\hm} = 0 
\end{equation}
and substituting again \eqn{frame} and \eqn{B} into \eqn{vary} we get
the equations of motion for $y^{\hm}$ 
\begin{equation}
\Box y^{\hm} + \eta^{IJ} \Gamma^{\hm}_{\ \hn\hr}\partial_I y^{\hn}
\partial_J y^{\hr} + \eta^{IJ} \partial_I y^{\hm}
\frac{\partial_J \rho}{\rho} = 0\ ,
\end{equation}
where the $\Gamma$'s are the Christoffel symbols of the seven--manifold metric
\begin{equation}
g_{\hm\hn}(y)={\cal B}^{\ha}_{\hm}(y){\cal B}^{\hb}_{\hn}(y)\eta_{\ha\hb}\,.
\end{equation}

It is interesting now to notice that, with the choices \eqn{P1}, \eqn{P2} and 
\eqn{frame}, these equations admit static solutions ($\rho = const$, 
$y^{\hm} = const$) if and only if 
$$
\q = 1.
$$
This fact selects the membrane action to be the one with $\q = 1$ and yields 
the recipe to fix the world--volume diffeomorphisms in a way consistent with 
$\kappa$-symmetry gauge--fixing.

\section{The supermembrane on the $AdS_4 \times S^7$ background}
\label{quattro}
Having constructed the $\kappa$--supersymmetric action \eqn{action}, 
in order
to continue our programme we have to
specialize it  to the $AdS_4 \times \SS^7$ background.
To achieve this point the coordinates of the  $D = 11$ target 
superspace
have to be split in the anti de Sitter ones and in that of the
seven--sphere.
After the splitting we need to fix a physical gauge such that
eight of these bosonic coordinates and eight of the fermionic ones become
fields on the brane world-volume.
 
To this end we have to find an explicit parametrization of the
vielbeins as functions of these fields and to fix the
three--dimensional diffeomorphisms and the $\kappa$--symmetry.
Usually one has to deal with very complex objects because the 32 
fermionic
coordinates of the $D = 11$ space mix with the bosonic ones in
complicated expressions.
This kind of analysis, though straightforward in line of principle,
is  very difficult to perform in practice.
 
A nice way to overcome this obstacle is to use the {\it Supersolvable}
parametrization of the vielbeins and the three--form  field.
This parametrization is perfectly equivalent to an a priori gauge--fixing of
$\kappa$--symmetry and allows to half the fermionic coordinates
(eight on the mass shell), simplifying the 
expressions one has to deal with.
 
As emphasized, the great technical advantage is 
that this fixes $\kappa$--symmetry a priori.
It implies that one does not have to calculate first long and 
complex expressions which one later gauge fixes, but one works with compact
formulae from the very start.

Let us then perform the $AdS_4\times\SS^7$ spontaneous compactification of
eleven--dimensional supergravity in a parametrization independent form
and find the superconformal gauge--fixed action for the probe membrane on such
a background.
\subsection{The $AdS_4 \times S^7$ splitting}
We already gave the constraints on the curvatures and torsions of the
$D = 11$ space in \eqn{3}.
From these and the solution of the Bianchi identities,  
it follows the dynamics of the fields described by their
equations of motion
\begin{eqnarray}
\Gamma^{\ua}_{\uaa\uab} T^{\uaa}_{\ua \ub} &=& 0, \nonumber \\
\label{eom}
R_{\ua\ub} - \frac{1}{2} \eta_{\ua\ub} R &=& - 288 \cdot 4! \left(
H_{\ua \uc_1 \uc_2 \uc_3} H_{\ub}{}^{\uc_1 \uc_2 \uc_3} - \frac{1}{8}
\eta_{\ua\ub} \, H_{\uc_1 \ldots \uc_4} H^{\uc_1 \ldots \uc_4}\right), \\
D_{\ud} H^{\ud}{}_{\ua\ub\uc} &=& - \frac{1}{4} \e_{\ua\ub\uc\ud_1 
\ldots
\ud_4 \ue_1 \ldots \ue_4} H^{\ud_1 \ldots \ud_4}  H^{\ue_1 \ldots 
\ue_4}. \nonumber
\end{eqnarray}
To find a consistent solution of these equations which parametrizes
$AdS_4 \times \SS^7$ all the fields and quantities are to be split
into four and seven dimensional ones and the antisymmetric tensor field 
$B$
has to satisfy the Freund--Rubin condition.
Following \cite{CreJul} the $D = 11$ ${\g}^{\ua}$ matrices can be 
expressed
in terms of the four--dimensional ones
$\g^a$ and the seven--dimensional $\tau^{\ha}$ as follows
\begin{equation}
{\g}^{\ua} = ( \unity\otimes \g^a, \tau^{\ha} \otimes \g^5)
\label{orfan14}
\end{equation}
and the charge conjugation matrix can be expressed in terms of the 
four and 
seven dimensional ones:
\begin{equation}
C = C_7 \otimes C_4 = \unity \otimes \g^0.
\end{equation}
This yields the splitting formulae for the eleven dimensional 
$\G^{\ua}$ matrices. 
The bosonic eleven--dimensional fields are relabeled as
\begin{equation}
E^{\ua} = (E^a, E^{\ha}), \quad {\omega_{\ua}}^{\ub} =
( {\omega_{a}}^{b}, {\omega_{a}}^{\hb},  {\omega_{\ha}}^{\hb}),
\quad B = B.
\label{orfan17}
\end{equation}
 
The Freund Rubin solution of eleven dimensional supergravity
can be obtained giving an expectation value to the field--strength
of the three--form  field
\begin{equation}
H_{a_1 \ldots a_4} = \frac{e}{4!} \e_{a_1 \ldots a_4},
\label{orfan18}
\end{equation}
and imposing
\begin{eqnarray}
\omega_{a\ha} &=& 0, \\
T_{\ua\ub}{}^{\uaa} &=& 0,
\label{orfan19}
\end{eqnarray}
which means the Lorentz connection factorizes and that 
the eleven--dimensional gravitino has vanishing vev.
 
From this, and \eqn{eom}, the Einstein equations become 
that
of an $AdS$ space and an Einstein seven--manifold
\begin{eqnarray}
R_{ab} &=& 48 e^2 \eta_{ab}, \label{orfan20a}\\
R_{\ha\hb} &=& -24 e^2 \delta_{\ha\hb},
\label{orfan20b}
\end{eqnarray}
of radius two times that of the anti de Sitter space.
In particular we can choose any homogeneous coset $G/H$.
The case of the seven sphere we are considering in this paper,
corresponds to the choice that leads to a
maximal number of preserved supersymmetries in anti de Sitter space 
(N=8).
This choice corresponds to the following Riemann tensors
\begin{eqnarray}
R_{ab}{}^{cd} &=& 32 e^2 \delta^{cd}_{ab}, \\
R_{\ha\hb}{}^{\hc\hd} &=& -8 e^2 \delta^{\hc\hd}_{\ha\hb}.
\end{eqnarray}
Since the Gaussian curvature of any sphere is $K = 1/R^2$ and the 
curvature
scalar is proportional to $K$, we can now relate the vev of the 
antisymmetric tensor field to the radii of the $AdS$ and $\SS^7$ 
spaces.
This relation is given by
\begin{equation}
\label{eR}
e = \frac{1}{2R}.
\end{equation}
 
A natural way to split the eleven--dimensional fermions is simply as 
in standard dimensional reduction \cite{dauria,repo} to write
$$
E^{\uaa} = E^{(A \alpha)} = \Psi^{\a}_A,
$$
which are the fermions living on
$$
{\cM}_{11} = \frac{OSp(8|4)}{SO(1,3) \times SO(7)}.
$$
There is however a rather more elegant way to proceed, which is to 
write 
instead
\begin{equation}
\label{splitferm}
E^{\uaa} = E^{(\hat{\a} \alpha)} = \sum_{A = 1}^8  \eta^{\hat{\a}}_A 
\otimes \psi_A^{\a},
\end{equation}
where $\psi_A^{\a}$ are Majorana $AdS_4$ fermions and  
$\eta_A$ are c--number $\SS^7$ Killing spinors,
functions only of $\rho$ and $y^{\ha}$.
This corresponds to the local decomposition
$$
{\cM}_{11} \approx \frac{OSp(8|4)}{SO(1,3) \times SO(8)} \times 
\frac{SO(8)}{SO(7)} = AdS_{(8|4)} \times \SS^7.
$$
The elegance of this approach lies in the simplicity of the resulting 
super--vielbeins and it
is indispensable for the generalisation to $G/H$ branes. 
To make \eqn{splitferm} consistent with the Bianchi identities it is 
necessary
to add fermion bilinears to the bosonic vielbein ${\cal B}^{\ha}$
and connection ${\cal B}^{\ha\hb}$ of $\SS^7$:
\begin{eqnarray}
E^{\ha} &=& {\cal B}^{\ha}(y) - \frac{1}{8} \, \eta_A \tau^{\ha} \eta_B 
{\cA}^{AB}(x,\theta), \\
\omega^{\ha\hb} &=& {\cal B}^{\ha\hb}(y) + \frac{e}{4} \,  
\eta_A \tau^{\ha\hb} \eta_B {\cA}^{AB}(x,\theta),
\end{eqnarray}
where ${\cA}_{AB}$ is the $SO(8)$ connection.

With the above choices the eleven dimensional constraints and 
Bianchi identities become relations on  the four--dimensional 
quantities.
The Ricci tensors for the $AdS_4$ space and for $\SS^7$ are then
\begin{eqnarray}
R^{ab} &=& -16 e^2 E^a \,\wedge\, E^b + 2 e \bpsi_A\, \wedge\, 
\g_{cd} \psi_A 
\e^{abcd},
\label{orfan21a}\\
R^{\ha\hb} &=& 4 e^2 E^{\ha}\, \wedge \, E^{\hb} + 2 e \bpsi_A \, 
\wedge 
\,\g^5 \psi_B \; \eta_A
\tau^{\ha\hb} \eta_B.
\label{orfan21}
\end{eqnarray}
The torsions and gravitinos satisfy
\begin{eqnarray}
\label{orfan22a}
D E^a  &=& \bpsi_A \,\wedge\, \g^a \psi_A, \\
D E^{\ha} &=& \eta_A \tau^{\ha} \eta_B \; \bpsi_A \,\wedge\, \g^5 
\psi_B, \label{orfan22b}\\
\rho_A \equiv D \psi_A &=& - 2e E^a \,\wedge\, \g_a \g^5 \psi_A - 
e {\cA}_{AC} \psi_C,
\label{orfan22c}
\end{eqnarray}
the $SO(8)$ connection satisfies
\begin{equation}
\label{SOA}
d {\cA}_{AB} + e {\cA}_{AC}\, \wedge \, {\cA}_{CB} = 
8 \bpsi_A \, \wedge \,\g_5 \psi_B,
\end{equation}
while the sphere spinors do indeed satisfy the Killing equation 
(cfr. equation \eqn{kilgspi})
\begin{equation}
D_{(Sph)} \eta_A = e {\cal B}^{\ha} \tau_{\ha} \eta_A,
\label{orfan23}
\end{equation}
where $D_{(Sph)} \equiv d + {\cal B}(y)$.
It is now  easy to
recognize that \eqn{orfan21a}--\eqn{SOA} are the Maurer Cartan
equations of the $OSp(8|4)$ supergroup \cite{dauria}.
In fact, eq.s \eqn{orfan21a},\eqn{orfan22a},
\eqn{orfan22c} and \eqn{SOA} can be rewritten as
\begin{eqnarray}
d \omega^{ab} + \omega^{ac} \, \wedge \,\omega_c{}^b + 16 e^2 E^a \, 
\wedge \,E^b &=&
4 e \bpsi_A \, \wedge \g^{ab} \g^5 \psi_A, \nonumber \\
\label{orfan25}
d E^a + \omega^a_c\, \wedge \, E^c &=& \bpsi_A \, \wedge \,\g^a 
\psi_A, \\
d \psi_A + \frac{1}{4} \omega^{ab}\, \wedge \, \g_{ab} \psi_A
+ e {\cA}_{AB} \, \wedge \,\psi_B &=& - 2 e 
E^a \, \wedge \,\g_a \g_5 \psi_A, \nonumber\\
d {\cA}_{AB} + e  {\cA}_{AC}\, \wedge \, {\cA}_{CB} &=& 8  \bpsi_A \, 
\wedge \,\g_5 \psi_B, \nonumber
\end{eqnarray}
which are the desired Maurer--Cartan equations of the $OSp(8|4)$
algebra given in standard form.
Indeed an element of the $OSp(8|4)$ superalgebra   can be defined
as a graded matrix
\begin{equation}
\mu = \left(
\begin{tabular}{c|c}
$ \displaystyle \frac{1}{4} \omega^{ab} \g_{ab} + 2e \, E^a \g_5 \g_a$ &
$-8 e \g_5 \psi_B$ \\\hline
$\phantom{\stackrel{A}{A}} \bpsi_A\phantom{\stackrel{A}{A}} $ &
$e {\cal A}_{AB}$
\end{tabular}
\right),
\label{orfan26}
\end{equation}
preserving the ortosymplectic matrix\footnote{Condition 
\eqn{symcondo} defines the complex orthosymplectic
superalgebra. In order to get the appropriate real section one usually
adds a pseudounitarity condition, namely:
\begin{equation}
 \label{pseucondo}
 H \mu +  \mu^\dagger H = 0.
\end{equation}
 This is the analogue of defining the bosonic anti de Sitter group
$SO(2,3)$ through the isomorphism $SO(2,3) \sim Usp(2,2)/\ZZ_2$ where
$Usp(2,2)\, \equiv \, Sp(4,\IC) \, \cap \, SU(2,2)$. However, we can
also define $SO(2,3)$ through the alternative isomorphism $SO(2,3) 
\sim
Sp(4,\IR)/\ZZ_2$. In this case it just suffices to consider real 
symplectic
matrices, removing the bosonic analogue of the second condition
\eqn{pseucondo}. Such a situation arises in the Majorana
representation of gamma matrices. Here the spinorial representation
which is symplectic is also real and naturally realizes the
isomorphism $SO(2,3) \sim Sp(4,\IR)/\ZZ_2$. Obviously all this
carries over to the super Lie algebra case. Choosing the Majorana
representation of gamma matrices it suffices to define the
orthosymplectic group as the set of {\it real } graded matrices
satisfying condition \eqn{symcondo} and discard condition 
\eqn{pseucondo}.
This is what we do here with our chosen basis of gamma matrices.}:
\begin{equation}
\Omega = \left( \bet{c|c} ${C} \g^5$ & $0$ \\\hline $0$ & $8e$
\eet \right),
\label{orfan27}
\end{equation}
i.e.
\begin{equation}
  \Omega \mu + {}^t \mu \Omega = 0.
  \label{symcondo}
\end{equation}
The Maurer Cartan equations \eqn{orfan25} can be
retrieved by writing $ d \mu + \mu\wedge\mu = 0$.
 
Now we come to the main point in the construction of the solution.
The curvature of the three--form $B$ can be expressed in terms of
$AdS_4 \times \SS^7$ quantities as follows:
\begin{eqnarray}
H &=& \frac{e}{4! \cdot 4!} E^a \, \wedge \,\ldots \, \wedge \,E^d 
\e_{d \ldots a} + \frac{1}{4! 4!}\, E^a \, \wedge \,E^b \, \wedge \, 
\bpsi_A \, \wedge 
\,\g_{ab} \psi_A + \nonumber \\
 &-&  \frac{2}{4! 4!} E^a \, \wedge \,E^{\hb} \, \wedge \,
\eta_A \tau_{\hb} \eta_B \; \bpsi_A \, \wedge \,
\g_5 \g_a \psi_B + \nonumber \\
 &-& \frac{1}{4!  4!}\, E^{\ha}\, \wedge \, E^{\hb} \, \wedge \,
\eta_A \tau_{\ha\hb} \eta_B \;
\bpsi_A \, \wedge \,\psi_B,
\label{orfan28}
\end{eqnarray}
and, from the curvature definition $dB = H$, it should be deduced the
parametrization of the three--superform $B$, but this can be done only
if one uses a specific parametrization in terms of the coordinates.
It is then useful to look at the supersolvable parametrization of our space.
\subsection{The supersolvable algebra and the $\kappa$--symmetry
gauge--fixing}
 
The $AdS$ superspace is defined as the following coset
$$
AdS^{(8|4)} = \frac{OSp(8|4)}{SO(1,3) \otimes SO(8)}
$$
and it is spanned by the four coordinates of the $AdS_4$ manifold
and by eight Majorana spinors (i.e. they have 32 real components)
parametrizing the fermionic generators $Q_{\a}^A$ of the superalgebra.
 
It has already been shown that the $AdS$ manifold admits a suitable
description in terms of a four dimensional solvable Lie algebra
$Solv$ \cite{GsuH}.
The problem which we deal with is that of finding a supersolvable
description of the superspace $AdS_{(8|4)}$.
It turns out that a solvable superalgebra $SSolv$ containing $Solv$
can be found inside $OSp(8|4)$, the only price one has to pay being a 
suitable
projection of the fermionic generators $Q_{\a}^A \to Q_{\a}^{\prime A}
= {\cal P} Q_{\a}^A$.
 
Just as the solvable description of $AdS_4$ allows to define the 
coordinates
on the brane it will be seen that the supersolvable description
of superspace yields the definition of the fermions living on the
brane as the result of the equivalence of the projection operator
${\cal P}$ on the target fermionic coordinates, and the 
$\kappa$--symmetry
projection operator.
 
It is perhaps worth pointing out that the supersolvable Lie algebra 
admits a very
natural physical interpretation. The starting point is the 
decomposition of the
$Osp(8|4)$ algebra of $AdS_{(8|4)}$ isometries in terms
of the superconformal algebra of the three dimensional worldvolume 
theory
of the brane.
 
We  can decompose the $SO(2,3)$ Lie algebra of invariances of $AdS$ 
in terms
of a 3 dimensional  $SO(1,2)$ sub-algebra $\lbrace L_{\perp}, L_{\pm} 
\rbrace$,
which is just the algebra of Lorentz rotations in the brane.
Of the six remaining step operators of $SO(2,3)$, three are
interpreted as worldvolume translations
$ \lbrace \tau_{\perp}, \tau_{\pm} \rbrace$,
and the other three are the conformal boosts $\lbrace
\sigma_{\perp}, \sigma_{\pm}
\rbrace$.
Of the two Cartan generators of  $SO(2,3)$, one goes into $SO(1,2)$,
while its orthogonal complement is the generator of dilatations $D$.
 
The fermions split as  ${\bf 4}\otimes{\bf 8}_V=({\bf 2}\oplus{\bf 
2})\otimes{\bf 8}_V$.
Physically each of the eight space-time supercharges, four--component 
$D=4$
Majorana spinors, split into eight worldvolume supercharges, 
two--component
$D=3$ Majorana spinors, and eight matching worldvolume superconformal 
generators.
 
This decomposition is illustrated in the familiar ``Union Jack'' 
root diagram of $C_2$, which is the complexification of $SO(2,3)$ 
shown in figure (1). 
The fermionic supercharges form
a square weight diagram within this figure, and the supertranslation 
algebra
is then simply that the anticommutator of two fermions is given by 
vector
addition of the corresponding weights in the diagram. The diagram can 
in fact
be seen as a projection of the full $Osp(8|4)$ root diagram, since 
the $SO(8)$
roots lie on a perpendicular hyperplane, and so on this diagram they 
would be at the
centre.
 
\iffigs
\begin{figure}
\epsfxsize=8cm \epsfysize=2.0in  
\centerline{\epsfbox{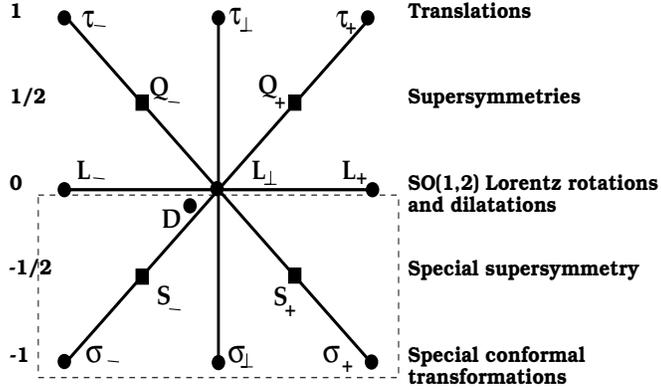}}
\caption{ {\small The root diagram of $SO(2,3)$. The bosonic weights 
are represented
by circles, and the fermionic weights by squares. The dilatation 
charge of
horizontal planes in the diagram are on the left, while the 
worldvolume
theory interpretation of the planes of generators are labelled to the 
right.
The supersolvable algebra is the boxed subalgebra. }}
\unitlength=1mm
\end{figure}
\fi
In figure (1) the decomposition into the algebra of the 
worldvolume superconformal
algebra is simply the decomposition into horizontal planes. A natural 
way to characterize
these planes is by their eigenvalues under $D$, i.e. by their 
dilatation charge. These charges
are given by the numbers on the left of the figure. More formally, 
this charge provides
a $\hbox{\msbm Z}_5$ grading to $Osp(8|4)$, a preserved charge under 
the super Lie bracket. This is
precisely the grading required to define the supersolvable Lie 
algebra of Appendix C,
namely the solvable algebra of $Osp(8|4)$ is just the sub-algebra 
obtained by restricting $Osp(8|4)$
to negative grading complemented by $D$ itself. Its weight diagram is 
boxed in
the diagram. It can also be seen  that in our explicit representation 
of gamma
matrices, this corresponds to restricting to triangular 
matrices.
The generators of this supersolvable algebra are in fact a suitable 
choice of generators
for the super coset space $AdS_{(8|4)}$. Their attraction comes from 
the fact that they are
easily exponentiated since the power series expansion of the 
exponential only contains a
finite number of terms, and so explicit expressions for the 
supervielbeins are easily
obtained.
 
The choice of a gauge fixing condition is a subtle point.
This condition must indeed be compatible with the classical solution 
of the brane--wave equations of motion chosen as the vacuum around 
which the perturbative theory is developed.

As it has been pointed out in \cite{PST}, to have the static 
solutions needed to perform the correct expansion around the boundary 
configurations, the Grassmann coordinates one has to project away 
gauge--fixing the $\kappa$--symmetry are those parallel to the 
$\kappa$--symmetry projector.
Since this same projector leaves the $Q$'s invariant (which are then 
recognised as the world--volume preserved supersymmetries) the proper 
choice to parametrize the super--$AdS$ space and obtain a $SSolv$ 
algebra is to use the generators $\{ S_{\pm}, \s_{\pm}, \s_{\perp}, 
D\}$.
 
More details about the supersolvable algebra and the parametrization 
of
the Super$AdS$ space can be found in the Appendix C.
 
We give here for the metric \eqn{nearhor} 
the parametrizations of the vielbeins in
terms of the four solvable coordinates ($\rho$, $t$, $w$, $x$)
and the eight four--dimensional fermions
($\theta_{\a}^A$):
\begin{eqnarray}
E^0 &=& - \rho dt - 2e \rho \bth^A \g^0 d \th^A, \nonumber \\
E^1 &=& \rho dw - 2e \rho \bth^A \g^1 d \th^A, \nonumber \\
E^2 &=& \frac{R}{2} \frac{1}{\rho} d\rho,  \label{param1} \\
E^3 &=& \rho dx - 2e \rho \bth^A \g^3 d \th^A, \nonumber
\end{eqnarray}
and
\begin{equation}
\psi^A = \sqrt{2e\rho}
\; \left( \begin{tabular}{c} $0$ \\ $0$ \\ $d\theta_1^A$ \\
$d \theta_2^A$ \end{tabular} \right),
\end{equation}
where $\displaystyle \th^A = \frac{1-\g^5\g^2}{2} \theta^A$ and
$\bth^A = \th^A \g^0$.
It can also be found that the $SO(8)$ connection $\cA$, in this 
parametrization, is identically zero:
\begin{equation}
{\cal A}_{AB} = 0.
\label{orfan11}
\end{equation}
To complete the parametrization of the target superspace one has
to define also the vielbeins of the seven--sphere.
Calling $y^{\ha}$ the seven coordinates of the sphere, the
parametrization we adopt is the following
\begin{equation}
\label{param5}
E^{\ha} = {\cal B}^{\ha} = -R \delta ^{\ha}_{\ \hm} \frac{d y^{\hm}}{1 + y^2}.
\end{equation}
which is nothing else than the stereographic projection coordinates.
 
The last thing to find is the parametrization of the Wess--Zumino 
term.
This means solving $H = dB$ in this background.
As we said in the last section this calculation can be done once obtained an
explicit parametrization.
So, with the parametrizations \eqn{param1}--\eqn{param5} given above, the 
$B$ field has the form
\begin{equation}
B = \frac{1}{4! \cdot 4!}\, E^i\, \wedge \, E^j\, \wedge \, E^k 
\,\frac{\e_{kji}}{3}\, - \, \frac{1}{2e} \, \frac{1}{4!4!} \,
E^{\ha} \, \wedge \, \eta_A \tau_{\ha} \eta_B \;\bpsi_A \, \wedge 
\,\psi_B.
\label{orfan29}
\end{equation}
We are now in position to obtain the complete action of the supermembrane
on an $AdS_4 \times \SS^7$ background.
The only missing point is the definition of the brane coordinates,
i.e. the choice of a physical gauge.
\subsection{The superconformal action in second order ``Nambu-Goto'' formalism}
The $\kappa$--symmetry has already been fixed using the solvable
parametrization.
We fix the three--dimensional world--volume diffeomorphisms 
imposing the static gauge choice.
As a consequence of what has been said in section \eqn{boseeqn}, 
to obtain static solutions we have to identify 
\begin{equation}
\xi^I \equiv (-t,w, x),
\label{orfan30}
\end{equation}
where $I=0,1,2$, is the curved index of the brane.
Once identified these coordinates, the form of the $\Pi$ fields
can be deduced from the vielbein parametrizations.
These can actually be projected along the basis of the brane
cotangent space:
\begin{equation}
E^{\ua} = e^i \Pi_i{}^{\ua} = d \xi^I \Pi_I{}^{\ua}.
\label{orfan31}
\end{equation}
The brane metric can then be deduced from the embedding equations 
using the above parametrizations
\begin{eqnarray}
h_{IJ}(\xi) &=& \Pi_I{}^{\ua}\Pi_J{}^{\ub}\eta_{\ua\ub} = 
\frac{1}{16 e^2} \frac{1}{\rho^2}
\partial_I \rho \partial_J \rho + \frac{1}{4 e^2 \, (1+y^2)^2} 
\partial_I 
y^{\ha} \partial_J y_{\ha} + \nonumber \\
\label{metrica}
&+& \rho^2 \left(\eta_{IJ} - 4 e \bth^A \g^i \partial_{(J} \th^A 
\delta_{i\,I)} + 4 e^2 \bth^A \g^i \partial_I \th^A \ \bth^B \g_i
\partial_J \th^B \right).
\end{eqnarray}
In second order formalism and with the embedded metric \eqn{metrica},
the action of the membrane on the $AdS_4 \times \SS^7$ background is 
now
\begin{equation}
\label{confaction}
S = 2 \int \sqrt{-\mbox{det}(h_{IJ})} \, d^3\xi + 4!4! \int B.
\end{equation}
The expression for $B$ is given by
\begin{eqnarray}
B &=& \frac{d^3 \xi}{4!4!}\left[ \e^{IJK} \rho^3 (\delta_I{}^i - 
2e \bth^A \g^i \partial _I \th^A)
(\delta_J{}^j - 2e \bth^A \g^j \partial_J \th^A)
(\delta_K{}^k - 2e \bth^A \g^k \partial _K \th^A)
\frac{\e_{ijk}}{3} + \right. \nonumber \\
&-& \left. \frac{1}{2 e (1+y^2)} \e^{IJK} \partial _I y^{\ha} \;  
\eta_A \tau_{\ha} \eta_B
\; \rho \partial _J \bth^A \partial _K \th^B \right] .
\label{orfan35}
\end{eqnarray}
 
It should be noticed that the field theory we have derived in this 
way has non trivial interactions  and is highly non--linear.
 
An important feature displayed by this action is its invariance
under conformal transformations.
As already stressed in the introduction and before in this section,
the super--$AdS$ group acts on the membrane as the group of
superconformal transformations.
This action is non--linearly realised \cite{renatoine1,kallosh1} and 
this partly
explains the complicated expression of \eqn{metrica} and 
\eqn{orfan35}.
 
We have completed here the programme started in 
\cite{renatoine1,renatoine2,renatoine3,kallosh1}, where the 
authors presented the theory restricted to only radial fluctuations 
or the free--field limit in the purely bosonic sector.
In fact, keeping the $y$ and $\theta$ fixed and reducing only 
to the radial fluctuations the action takes the form
\begin{equation}
S = 2 \left\lmoustache{
\rho^3 \left[ \sqrt{1 + \frac{R^2}{4} \frac{\partial_I \rho 
\; \partial^I \rho}{\rho^4}} - 1 \right] \, d^3 \xi}\right.,
\end{equation}
from which we can recover the action (15) presented in 
\cite{kallosh1}, 
setting 
\begin{equation}
\rho = \frac{\phi}{R^2}, 
\end{equation}
where $w = \frac{1}{2}$ and $p =2$.
\par
If one properly identifies the radius $R$ as \cite{maldapasto}
$$
R = l_p(2^5\pi^2N)^{1/6},
$$
where $l_p$ is the eleven--dimensional Planck length and
$N$ is the number of $M2$--branes, \eqn{confaction} can be
interpreted as the action of one probe membrane in the background
generated by the other $N-1$ branes.
 
The explicit form of the non--linear realisation of the 
superconformal transformations on the world--volume
is much simplified by the
observation that although there are 6 superconformal generators,
namely $\lbrace D,\sigma_{\perp},\sigma_{\pm},S_{\pm}\rbrace$ 
(see figure (2)),
one  needs to verify invariance only for two,
namely dilatations and special conformal inversion.
This is so because we can explicitly construct the operator that
implements the Weyl reflection in the horizontal $0$--grading plane
(see figures (1) and (2)).
\iffigs
\begin{figure}
\epsfxsize=6cm \epsfysize=4cm
\centerline{\epsfbox{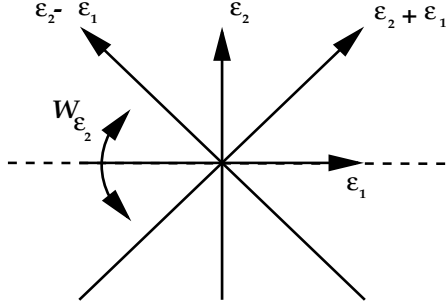}}
\caption{ {\small The root diagram of $SO(2,3)$, with simple roots 
$\epsilon_1$
and $\epsilon_2$. The Weyl reflection $W_{\epsilon_2}$ is shown in 
the picture. }}
\unitlength=1mm
\end{figure}
\fi
Thus any superconformal generator in the lower part of the weight 
diagram can be
constructed as a worldvolume Poincar\'{e} generator conjugated 
through the
Weyl reflection $W_{\epsilon_2}$.
Given that our worldvolume theory is Poincar\'{e} invariant, all that 
remains
to be checked is dilatation invariance and invariance with respect to
the finite group element $W_{\epsilon_2}$.
We have:
\begin{equation}
W_{\epsilon_2} \ = \ 
\exp{\left[\pi(E_{+\epsilon_2}-E_{-\epsilon_2})\right]}
 \ = \ \exp{\left[\pi(\tau_{\perp}-\sigma_{\perp})\right]}.
\end{equation}
Thus we only need the transformation induced by 
$\tau_{\perp}-\sigma_{\perp} \equiv K_3$ in order
to be able to write any superconformal transformation. The action of 
the $Osp(8|4)$
generators on the coset space $AdS_{(8|4)}$ is simply given by the 
corresponding
Killing vector.
The transformation of the $\rho$ field, the coordinates and the
spinor fields under dilatations is
\begin{eqnarray}
\delta \rho &=& \rho, \nonumber \\
\delta \xi^I &=&  - \xi^I, \label{dilat} \\
\delta \theta^A_{\a} &=& - \frac{1}{2}\theta^A_{\a}, \nonumber
\end{eqnarray}
while their transformation under the Weyl rotation generator $K_3$ is
\begin{eqnarray}
\delta \rho &=& \rho \xi^2, \nonumber\\
\delta \xi^0 &=& - \xi^0 \xi^2, \nonumber\\
\delta \xi^1 &=& - \xi^1 \xi^2, \nonumber\\
\label{K3}
\delta \xi^2 &=& - \frac{1}{2} (1 + (\xi^2)^2) + \frac{1}{2}
((\xi^1)^2 - (\xi^0)^2)-\frac{1}{2\rho^2}
-  \frac{e^2}{2} \theta_1^A \theta_2^A \ \theta_1^B \theta_2^B, \\
\delta \theta^A_1 &=& - \frac{1}{2} \xi^2 \theta^A_1, \nonumber\\
\delta \theta^A_2 &=& - \frac{1}{2} \xi^2 \theta^A_2 . \nonumber
\end{eqnarray}
An important thing to note here is that the transformation of the 
fields
$f(\xi)$ dependent on the brane coordinates $\xi^I$ is their {\it 
complete}
variation for the conformal transformations, i.e. $\delta f =
f^{\prime} (\xi^{\prime}) - f(\xi)$, in that \eqn{dilat}--\eqn{K3}
express the functional variation plus the $\xi$
dependent transformation of the fields.
This observation leads then to the following identity for the
variation of the derivatives of the fields
\begin{equation}
\delta (\partial_I f(\xi)) = \partial_I \delta f - \partial_J f 
\partial_I \delta \xi^J,
\label{orfan38}
\end{equation}
which is very useful for the verification of invariance  of the action
\eqn{confaction} under the transformations \eqn{dilat} and \eqn{K3}.
\section{The singleton action from the supermembrane}
\label{cinque}
Starting from the action \eqn{confaction} 
we can now recover the singleton field theory expanding 
the transverse coordinates  around the
classical solution of the brane--wave equations provided by
\begin{equation}
 \xi^I \equiv (-t,w,x), \quad \partial _I y^{\hm} = 0,
\quad \theta^{\a}_A = 0,
\label{orfan39}
\end{equation}
and 
\begin{equation}
\rho = \bar{\rho} = \mbox{const} \quad : \quad 
\rho \to \infty \quad \hbox{ or } \quad \rho \to 0.
\end{equation}
This means choosing a physical gauge, a point on the
seven--sphere and then taking the limit to the boundary.
We would like to stress here that the proper conformally invariant 
boundary of the $AdS$ space is given by the three--dimensional 
Minkowski space at $\rho \to \infty$ with some points at its infinity added. 
We will find the singleton field theory on such 
boundary.
But it can be seen that the $AdS$ isometries act on the horizon such 
that it can be constructed a conformal invariant theory expanding 
around $\rho = 0\ $.  
Thus, we are going to find that the singleton field theory 
describes the centre of mass degrees of freedom of the $M2$--brane as 
a solution of the eleven--dimensional supergravity 
equations of motion.

We define the quantum expansion (small fluctuations) around
the classical solution using the normal coordinate expansion 
\cite{bkgr}:
\begin{equation}
X^{M} = x^{M} + \a^{\prime \frac{3}{2}} \f^{M}\ .
\label{orfan40}
\end{equation}
where $x^M$ is the background value for $X^M$ (the $11 + 32$ superspace 
coordinates), $\f$ is the normal 
coordinate (quantum field) and $\a^{\prime}$ is related to 
the membrane tension (now $M = (m,\mu)$). 
 
For the sake of simplicity we take the $y^{\hm} =  0$ point on the 
sphere.
Thus, the expansion formulae for the coordinates are
\begin{eqnarray}
\rho &=& \bar{\rho} + {\alpha'}^{\frac{3}{2}} \tilde\rho\,, 
\nonumber\\
\label{fluttua}
y^{\hm} &=& {\alpha'}^{\frac{3}{2}} \tilde y^{\hm} \,, \\
\th^A{}_{\a} &=& {\alpha'}^{\frac{3}{2}} \Theta^A_{\a}. \nonumber 
\end{eqnarray}
$\bar{\rho}$ is different from zero, but constant.
 
Applying the normal coordinate expansion formulae as they are given in
section A.1 to the supermembrane action, we can expand this
as a power series in $\a^\prime$
\begin{equation}
\label{orfan41}
{\cal L} = \sum_{n = 0}^{\infty} \a^{\prime}{}^{\frac{3(n-2)}{2}}
{\cal L}_{(n)}
\end{equation}
The result in terms of the fluctuations \eqn{fluttua} is that the 
vacuum membrane graph and the tadpole term are exactly zero:
\begin{equation}
\label{orfan42}
{\cal L}_{(0)} = 0 = {\cal L}_{(1)}
\end{equation}
and we should recover the singleton action from the order 1 term
\begin{eqnarray}
{\cal L}_{(2)} &=& \frac{1}{16 e^2\bar{\rho}} \eta^{IJ} 
\partial_I \tilde\rho \,
\partial_J \tilde\rho + \frac{\bar{\rho}}{4 e^2} \eta^{IJ} 
\partial_I \tilde y^{\hm} \,
\partial_J \tilde y^{\hn} \delta_{\hm\hn} - 
4 e \, \bar{\rho}^3 \; \bar{\Theta}^A \hat\sigma^i
\partial_I \Theta^A \delta_i^I.
\label{orfan47}
\end{eqnarray}
 
As it can be easily seen, taking the boundary limit, some terms in 
the action \eqn{orfan47} vanish and some other diverge.
This implies that, in order to keep it finite, we have to rescale
these fluctuations with some power of $\bar{\rho}$.
This does not provide us the singleton action yet.
Naively speaking in fact, one can see that all the other terms 
disappear
as we go to the boundary.
But we did not take into account the symmetry transformations.
We will actually see that also these transformations diverge on the 
same limit and that the right rescaling of the fluctuations 
are the ones which let us retrieve the singleton action from 
\eqn{orfan47}.
 
Let us analyse then the symmetry of \eqn{orfan47}.
Since the action we obtained has $\kappa$--symmetry fixed, the 
supersymmetry transformations should preserve this gauge fixing.
To this end one has to accompany any SUSY transformation with a 
$\kappa$--symmetry transformation such that one does not move the 
chosen
configuration.
 
Following \cite{KalloshN}, we have fixed $\kappa$--symmetry by 
imposing
\begin{equation}
\label{kfixing}
(1 + \bG) \theta = 0\ .
\end{equation}
Calling $\chi = \left( \begin{array}{c} \chi_L \\ \chi_R \end{array} 
\right)$,
\eqn{kfixing} can be read as $\theta_R = 0$.
 
In the $\g$--matrices basis chosen in the Appendix A, the generic
value of $\bG$ has the following block structure:
\begin{equation}
\label{proiet}
\bG = \left( \begin{array}{cc} -ACA^{-1} & A \\
  (1-C^2) A^{-1} & C
\end{array} \right),
\end{equation}
enforced by the condition $\bG^2=1$.
For a SUSY plus  $\kappa$--symmetry variation, the variation of the 
fermions is
\begin{equation}
\delta \theta = \e + \frac{(1+\bG)}{2} \kappa,
\end{equation}
which, from \eqn{proiet}, can be written as
\begin{eqnarray}
\delta \theta_L = \e_L + \frac{A}{2} \kappa_R, \nonumber \\
\delta \theta_R = \e_R + \frac{1+C}{2} \kappa_R.
\end{eqnarray}
To preserve the \eqn{kfixing} gauge fixing, one has to impose
$\delta \theta_R =0$.
Therefore, the compensating $\kappa$--symmetry transformations
has parameter $\kappa_R = -2(1+C)^{-1} \e_R$.
Thus, the complete SUSY transformation of the physical fermions is
\begin{equation}
\delta \theta_L = \e_L - A(1+C)^{-1} \e_R\ .
\end{equation}
 
Since we also fixed the world--volume diffeomorphisms imposing the 
static gauge \eqn{orfan39}, the total variation of 
$\theta_R$ as a field on the world--volume is
\begin{equation}
\label{tran1}
\delta \th^A = \e^A_- - A(1+C)^{-1} \e_+^A - \partial_I \th^A 
\; \delta_{\e +\kappa} \,x^I
\end{equation}
while the other field transformations are
\begin{eqnarray}
\label{tran2}
\delta \rho &=& \delta_{\e} \rho - \partial_I \rho \;  
\delta_{\e +\kappa} \, x^I \, ,  \\
\delta y^{\ha} &=& \delta_{\e} y^{\ha} - \partial_I y^{\ha} \;  
\delta_{\e +\kappa} \, x^I \, . \nonumber
\end{eqnarray}
To derive the explicit form of these transformations we have then to 
find the value of $A$ and $C$ on our background, while to derive the
transformation of the fluctuations one has also to expand the above
equations and identify the terms with the same powers of $\a^\prime$.
 
As it is known, the classical configuration specified by 
\eqn{orfan39} cannot 
preserve all the target space SUSY, but in the best case (like this)
it preserves the half.
 
Our choice of the vacuum \cite{fund} imposes that in order to 
preserve 
SUSY
\begin{equation}
\label{VEV}
\delta \psi = \delta \theta = 0.
\end{equation}
Since we projected the $\theta$ with the relation \eqn{kfixing}, 
condition
\eqn{VEV} translates into the fact that the residual SUSY are those 
transformations parametrized  by an $\e$ which satisfies
\begin{equation}
\tilde{D} \e = 0 \quad \hbox{and} \quad \frac{(1-\bG)}{2}\e = 0\ ,
\end{equation}
where $\tilde D$ is the supercovariant derivative and then implies
that $\e$ is a killing spinor.
Thus we are left with the transformations \eqn{tran1} and \eqn{tran2} 
in 
which we set $\e_L = 0$.
 
We want the SUSY transformations on the world--volume. We have then 
to take the SUSY transformations \eqn{tran1}--\eqn{tran2} 
and make the expansion \eqn{fluttua}.
It is straightforward to find that
\begin{eqnarray}
C &=& +1, \\
A &=& \a^{\prime\,\frac{3}{2}} \left[ \frac{\partial_I 
\tilde{y}^{\ha}}{2e \bar{\rho}} \; \tau_{\ha} - \frac{\partial_I 
\tilde{\rho}}{4 e \bar{\rho}^2}\right] \hat{\s}^I \hat{\s}^0,
\end{eqnarray}
and, matching orders in $\a^{\prime}$,
\begin{eqnarray}
\delta \Theta_-^A &=& \frac{1}{2} \left[ \frac{\partial_I 
\tilde{y}^{\ha}}{2e \bar{\rho}} \; \eta^A \tau_{\ha} \eta^B \; 
- \frac{\partial_I
\tilde{\rho}}{4 e \bar{\rho}^2} \; \delta^{AB} \right] 
\hat{\s}^I \hat{\s}^0 \e^B_+, \nonumber \\
\label{SUSY}
\delta \tilde{\rho} &=& - 8 e^2 \, \bar{\rho}^2 \; \bar{\e}_+^A 
\hat{\s}^0 \Theta_-^A, \\
\delta \tilde{y}^{\ha} &=& 4 e^2 \, \bar{\rho} \; 
\eta^A \tau^{\ha} \eta^B \; \bar{\e}^A_+ \hat{\s}^0 \Theta_-^B\, , 
\nonumber
\end{eqnarray}
where we have identified $i$ with $I$ since we are on flat space.
 
As we have already claimed, these transformations diverge as 
$\bar{\rho} \to 0$, but we can make them all finite if we rescale the 
fluctuations in the following way
\begin{equation}
\l = \bar{\rho}^{\frac{3}{2}} \Theta_+^A, \quad \tilde{P} = 
\frac{\tilde{\rho}}{\sqrt{\bar{\rho}}}, \quad \tilde{Y}^{\ha} = 
\sqrt{\bar{\rho}} \tilde{y}^{\ha}.
\end{equation}
 
In the \eqn{SUSY} transformations there appear the killing spinors 
on the sphere $\eta^A$.
These are functions of $y$ and, through these, of $\xi^I$.
When fixing the gauge and taking the expansion \eqn{fluttua}, the 
leading 
term in $\a^{\prime}$ for the $y$'s is zero and thus we can interpret 
$\eta^A \tau_{\ha} \eta^B$ as a simple numerical matrix.
 
Finally, it can be seen that if we define
\begin{equation}
Y^{\uA} \equiv \left\{\frac{\tilde{P}}{4}, \frac{\tilde{Y}}{2} 
\right\}
\qquad 
k \equiv \{-\delta^{AB}, \eta^A \tau^{\ha} \eta^{B} \}
\end{equation}
we can rewrite the singleton action as
\begin{equation}
\label{singleton}
{\cL} =  \eta^{IJ} \, \frac{1}{e^2} \, \partial_I Y^{\uA} \partial_J 
Y^{\uA} 
- 4 e \, \bl^A \hs^I \partial_I \l^A,
\end{equation}
with supersymmetry transformations given by
\begin{eqnarray}
\delta \l^A &=& \frac{1}{2e} \,  k^{A}_{\uA\dot{B}} \; \partial_I 
Y^{\uA}  
\; \hat{\s}^I \hat{\s}^0 \e^{\dot{B}}_+, \\
\delta Y^{\uA} &=& 2e^2 \, k^{\uA}_{\dot{B}C} \; \bar{\e}_+^{\dot{B}}
\hat{\s}^0 \l^C, 
\end{eqnarray}
where we reinterpreted the indices as if
$$
A \in \hbox{\bf 8}_S, \quad \dot{A} \in \hbox{\bf 8}_C, \quad
\uA \in \hbox{\bf 8}_V,
$$
and $ k^{\uA}_{\dot{B}C}$ is proportional to the triality matrices of 
$SO(8)$.
 
All the other superconformal transformations, under which 
\eqn{singleton} 
is invariant, can be retrieved applying the same method to the 
full transformations.
We have verified  that \eqn{singleton} is indeed invariant under 
dilatations, supersymmetry and Weyl reflections and thus on the whole 
algebra.

Since the action and the supersymmetry transformations are 
independent 
from $\bar{\rho}$, one should think that this action is 
superconformally invariant for any vev of $\rho$, but this is not the 
case.
If $\rho$ has a finite vev value, different from zero, there are 
$\a^\prime$ corrections to the action and the $K_3$ transformation is 
no more a symmetry of the theory since it depends explicitly on such 
vev $\bar{\rho}$.
 
\bigskip

\paragraph{Acknowledgements.}
\ We are grateful to A.Lerda for many useful discussions and to
G. Arcioni, F. Cordaro, L. Gualtieri for collaboration in the
early stages of this work.
We would also like to thank P. Pasti, D. Sorokin and M.Tonin for valuable 
comments and discussions about the proper choice of a consistent
$\kappa$--symmetry gauge fixing.

\bigskip

\section*{Appendix A: Notations and Conventions}
\setcounter{equation}{0}
\makeatletter
\@addtoreset{equation}{section}
\makeatother
\renewcommand{\theequation}{A.\arabic{equation}}
Latin letters are the indices of the bosonic coordinates, greek 
indices label the fermionic ones.
Letters from the beginning of the alphabet are flat indices while middle
alphabet letters label curved indices.
Underlined indices refer to the eleven dimensional coordinates,  $\ua,\um 
= 0, \ldots, 10$,
$\uaa,\uam = 1, \ldots 32$; normal indices span the $AdS$ space $a,m = 0, 
\ldots 3$,
$\a,\mu = 1, \ldots 4 $ and the hatted indices label the seven--sphere 
$\hbox{\msbm S}^7$, $\ha,\hm = 1, \ldots 7$,
$ \hat{\alpha},\hat{\mu} = 1, \ldots 8$.
The membrane worldvolume is spanned by three bosonic coordinates 
labelled by $I = 0,1,2$ or
$i = 0,1,2$, if curved or flat indices respectively.
We use the mostly plus metric, i.e.
\begin{equation}
\eta_{\ua\ub} = \hbox{diag}\{- + + + + + + + + + +\}, \qquad 
\eta_{ij} = \hbox{diag}\{-++\}.
\end{equation}
 
A p-form $\phi_p$ is defined by
\begin{equation}
\phi_p = \frac{1}{p!} E^{\ua_1} \wedge \ldots \wedge E^{\ua_p} 
\phi_{\ua_p \ldots \ua_1}; 
\end{equation}
the differential acts from the right
\begin{equation}
 d(A_p B_q) = A_p \, d B_q + (-1)^p dA_p \, B_q 
\end{equation}
and the Levi-Civita tensor is defined as $\e_{012} = +1$.

The {\bf eleven--dimensional} gamma matrices ${\g}^{\ua}$ 
are elements of the Dirac algebra
\begin{equation}
\{ {\g}^{\ua}, {\g}^{\ub} \} = 2 \eta^{\ua\ub}.
\end{equation}
Through the charge conjugation matrix $C$ we define the matrices
\begin{equation}
\begin{array}{rcl}
(\G^{\ua})^{\uaa}{}_{\uab} &\stackrel{\phantom{A}}{\equiv}& 
({\g}^{\ua})^{\uaa}{}_{\uab} \\
(\G^{\ua})_{\uaa\uab} &\stackrel{\phantom{A}}{\equiv}& 
 C_{\uaa \uag}({\g}^{\ua})^{\uag}{}_{\uab} \\
(\G^{\ua})^{\uaa\uab} &\stackrel{\phantom{A}}{\equiv}& 
 ({\g}^{\ua})^{\uaa}{}_{\uag} C^{\uag \uab} \\
(\G^{\ua})_{\uaa}{}^{\uab} &\stackrel{\phantom{A}}{\equiv}& 
  C_{\uaa \uag} ({\g}^{\ua})^{\uag}
{}_{\uad}  C^{\uad \uab}.
\end{array}
\end{equation}

Antisymmetrization $\G^{a_1 \ldots a_n} \equiv \G^{[a_1} \ldots
\G^{a_n]}$ is understood with unit weight.
The symmetric matrices are 
\begin{equation}
\G^{\ua}, \G^{\ua \ub}, \G^{\ua_1 \ldots \ua_5}, \G^{\ua_1 \ldots 
\ua_6},
\G^{\ua_1 \ldots \ua_9}, \G^{\ua_1 \ldots \ua_{10}}.
\end{equation}
The ciclic identity in eleven dimensions reads
\begin{equation}
(\Gamma^{\ua\ub})_{(\uaa\uab} 
(\Gamma_{\ub})_{\uag\underline{\delta})} = 0.
\end{equation}

The {\bf four--dimensional} gamma matrices $\g^a$ satisfy
the Dirac algebra
\begin{equation}
\{ \g^a, \g^b \} = 2 \eta^{ab}.
\end{equation}
The $\g^5$ is defined through the relation
\begin{equation}
\g^5 \equiv - \g^0 \g^1 \g^2 \g^3 = - \frac{1}{4!} \e_{abcd}  \g^a \g^b 
\g^c \g^d.
\end{equation}
Our (completely real) parametrization is given by
\begin{equation}
\label{matr1}
\g^0 = \left( \begin{tabular}{cc} $-i \sigma^2$ & 0 \\ 0 & 
$i\sigma^2$ \end{tabular} \right),
\qquad
\g^1 = \left( \begin{tabular}{cc} $- \sigma^3$ & 0 \\ 0 & $-\sigma^3$ 
\end{tabular} \right),
\end{equation}
$$
\g^2 = \left( \begin{tabular}{cc} $0$ & $-i \sigma^2$  \\ $i 
\sigma^2$ & $0$ \end{tabular} \right),
\qquad
\g^3 = \left( \begin{tabular}{cc} $\sigma^1$ & 0 \\ 0 & $\sigma^1$ 
\end{tabular} \right),
$$
and
\begin{equation}
\label{matr5}
\g^5 = \left( \begin{tabular}{cc} $0$ & $i \sigma^2$  \\ $i \sigma^2$ 
& $0$ \end{tabular} \right).
\end{equation}
The charge conjugation matrix is $C = \g^0$ and
\begin{equation}
\g^0 \g^a \g^0 = \g^a{}^{\dagger} = {}^t \g^a
\end{equation}

The {\bf seven--dimensional} gamma matrices  are $\tau^{\ha}$, and 
satisfy
\begin{equation}
\{ \tau^{\ha}, \tau^{\hb}\} = - 2 \delta^{\ha\hb}.
\end{equation}
The killing spinors on the sphere are $\eta_A^{\hat{\a}}$ (where $A = 
1,
\ldots 8$ is the index of the {\bf 8}${}_S$ of $SO(8)$), they are
completely real (i.e. $\bar{\eta} = {}^t\eta$) and satisfy the
identity
\begin{equation}
\eta_A^{\hat{\a}} \eta_B^{\hat{\a}} = \delta_{AB}
\end{equation}
Some useful $\tau$ identities are:
\begin{eqnarray}
\eta_A \tau^{\ha\hb} \eta_B \; \eta_C \tau_{\hb} \eta_D {\cA}^{AB} 
{\cA}^{CD}
&=& 4 \eta_A \tau^{\ha} \eta_B {\cA}^{AC} {\cA}^{CB} \\
(\eta_A \delta_{BC}) {\cA}^{AB} &=& \frac{1}{16} \tau_{\ha\hb}\eta_C 
\, \eta_A
\tau^{\ha\hb} \eta_B {\cA}^{AB} - \frac{1}{8} \tau_{\ha} \eta_C 
\eta_A 
\tau^{\ha} \eta_B {\cA}^{AB} \\
\eta_A\tau^{\ha\hc} \eta_B  \eta_C \tau^{\hc\hb} \eta_D 
{\cA}^{AB}{\cA}^{CD} 
&=& \frac{1}{2}  \eta_A \tau^{\ha} \eta_B  
\eta_C \tau^{\hb} \eta_D {\cA}^{AB} {\cA}^{CD}
- 4  \eta_A \tau^{\ha\hb} \eta_B {\cA}^{AC} {\cA}^{CB}  
\end{eqnarray}

The {\bf three--dimensional} gamma matrices   are
\begin{equation}
\hs^I \equiv \{ i \s^2,  \s^3, -\s^1 \},
\end{equation}
and satisfy
\begin{equation}
{}^t \hs^I = \hs^0 \hs^I \hs^0.
\end{equation}
We also write here two identities useful to verify the
$\kappa$--symmetry of the action:
\begin{eqnarray}
\label{6}
\bG \G_{ij} \e^{ijk} &=& 2 \sqrt{-g} g^{kj} \G_j,  \\
\bG \G_i &=& \frac{1}{2} \frac{\e^{ljk}}{\sqrt{-g}} \G_{jk} g_{il}.
\end{eqnarray}
\subsection*{A.1: The membrane action and normal coordinates}
To expand our membrane action around the classical solution 
\eqn{orfan39}
we make use of the normal coordinates  \cite{bkgr,Eisenhart}.
 
Though we do not enter the details, we give here some useful
formulae as reference.
We consider the normal coordinate expansion
\begin{equation}
X^M = x^M + {\alpha '}^{\frac{3}{2}} \phi^M(x) +{\cal O}({\alpha 
'}^{3})\,,
\end{equation}
where $x^M$ is the classical value of $X^M$ and $\phi^M$ is the normal
coordinate (quantum fluctuation).
Its derivative with respect to the world-volume indices $\xi^I$ is
\begin{equation}
\partial_I X^M = \partial_I x^M + {\alpha'}^{\frac{3}{2}} \nabla_I
\phi^M - \frac{1}{3} {\alpha'}^{3}  \partial_I x^S R^M{}_{PSQ} \phi^P 
\phi^Q + {\cal O}({\alpha '}^{\frac{9}{2}})\,.
\end{equation}
From this one can derive 
\begin{eqnarray}
g_{MN}\big(X^M\big) &=& g_{MN}(x)
- \frac{1}{3} {\alpha '}^{3}R_{M S_1 N S_2} \phi^{S_1} \phi^{S_2} + 
{\cal O}({\alpha '}^{\frac{9}{2}})\,,
\nonumber \\
B_{MNP}\big( X^M  \big) &=& B_{MNP}( x ) + {\alpha '}^{\frac{3}{2}} 
\phi^S 
\nabla_S B_{MNP}( x )   \\
& & + \frac{1}{2} {\alpha '}^{3} \Big( \nabla_{(S_1}
\nabla_{S_2)} B_{MNP} (x) + R^T{}_{S_1 S_2 [M} B_{NP]T}(x) \Big)
\phi^{S_1} \phi^{S_2}
+{\cal O}({\alpha '}^{\frac{9}{2}})\,, \label{Bexpansion}
\nonumber
\end{eqnarray}
and using these we find for the induced metric on the membrane world
volume 
\begin{eqnarray}
h_{IJ}\big( X \big) &=& h_{IJ}(x) + 2 {\alpha '}^{\frac{3}{2}}
\partial_{(I} x^M \nabla_{J)} \phi^N g_{MN} \nonumber\\ 
&& + {\alpha '}^3 \Big( g_{MN} \nabla_I \phi^M \nabla_J \phi^N - 
R_{MPNQ}
\partial_I x^M \partial_J x^N \phi^P \phi^Q \Big) + {\cal O}({\alpha 
'}^{\frac{9}{2}})
\end{eqnarray}
and for the inverse 
\begin{eqnarray}
h^{IJ}\big( X \big) &=& h^{IJ}(x) - 2 {\alpha
'}^{\frac{3}{2}} \, h^{IK} \, h^{JL} \, \partial_{(K} x^M \nabla_{L)} 
\phi^N
g_{MN} + {\cal O}({\alpha '}^{3}) \,.
\end{eqnarray}
From these one can also recover the following expression for the 
determinant of the metric:
\begin{eqnarray}
\sqrt{-h}(X) &=& \sqrt{-h}(x)\nonumber\\
             & & + {\alpha '}^{\frac{3}{2}}\sqrt{-h}(x) \,  h^{IJ}
g_{MN} \partial_{I} x^M \nabla_{J} \phi^N \nonumber\\
&& 
+ \frac{1}{2}{\alpha '}^3 \sqrt{-h}(x) \Big[
      h^{IJ} g_{S_1 S_2} \nabla_I \phi^{S_1} \nabla_J
             \phi^{S_2}
      -h^{IJ} R_{M S_1 N S_2} \partial_I x^M \partial_J
             x^N \phi^{S_1} \phi^{S_2} \nonumber\\
&&
+\nabla_I \phi^{S_1} \nabla_J
             \phi^{S_2} g_{M S_1} g_{N S_2}(\partial^I x^M \partial^J 
x^N - \partial^J
             x^M \partial^I x^N - h^{IJ} \partial_K x^M \partial^K x^N
             )
\Big] \nonumber\\
&& +{\cal O}({\alpha '}^{\frac{9}{2}})\,. \label{sqrthexpansion}
\end{eqnarray}
 
Filling these formulae in the action for the membrane,
\begin{equation}
S = \frac{2}{{\alpha'}^3} \int \sqrt{-h} d^3 \xi + 
\frac{4!4!}{{\alpha '}^3} \int B_{MNP}(X) \partial_I X^M
\partial_J X^N \partial_K X^P \epsilon^{IJK} d^3 \xi \,,
\label{singletonaction}
\end{equation}
and expanding it in powers of $\alpha '$,
\begin{equation}
{\cal L} = \sum_{n=0}^\infty {\alpha '}^{\frac{3(n-2)}{2}} {\cal 
L}_{n}\,,
\end{equation}
we get
\begin{eqnarray}
{\cal L}_0 &=&  2 \sqrt{-h} + 4!4! \, B_{MNP}\, \partial_I x^M
\partial_J x^N \partial_K x^P \epsilon^{IJK} \,, \nonumber
\\
{\cal L}_1 &=&  2 \sqrt{-h}\, h^{IJ}\, g_{MN}\, \partial_I x^M
\nabla_J \phi^N + 4!4!\phi^S H_{SMNP}\, \partial_I x^M 
\partial_J
x^N \partial_K x^P \epsilon^{IJK} \,,\nonumber\\ 
{\cal L}_2 &=& \sqrt{-h} \,
      h^{IJ} g_{MN}\, \nabla_I \phi^{M} \nabla_J
             \phi^{N}
      - \sqrt{-h} \, h^{IJ} R_{M P N Q}\, \partial_I x^M \partial_J
             x^N \phi^{P} \phi^{Q} \nonumber\\
&+& 
 \sqrt{-h} \, \nabla_I \phi^{P} \nabla_J
   \phi^{Q} g_{M P}\, g_{N Q}\, (\partial^I x^M \partial^J x^N - 
\partial^J
             x^M \partial^I x^N - h^{IJ} \partial_K x^M \partial^K x^N
             ) 
 \nonumber\\
&+& 4!4!  \frac{3}{2} \phi^S \nabla_I \phi^M H_{SMNP}\,
\partial_J x^N \partial_K x^P \epsilon^{IJK} \nonumber\\
&+&  4!4! \frac{1}{2} R^T{}_{QS[M}\, B_{NP]T}\, \phi^Q \phi^S 
\partial_I
x^M \partial_J x^N \partial_K x^P \epsilon^{IJK}\,,
\label{explagr}
\end{eqnarray}
which are the formulae used in the text to derive
\eqn{orfan42} and \eqn{orfan47}.
\section*{Appendix B: The boundary of the universe}
\setcounter{equation}{0}
\renewcommand{\theequation}{B.\arabic{equation}}
As we have already said in section two, the four--manifold we consider
as universe is ${\cal U} = AdS_4/\ZZ_2$.
It can be partly covered by the chart (\ref{eq:diffeo}), diffeomorphic
to the upper half $\{\rho>0\}$ of $\IR^4$ equipped with the metric:
\begin{equation} \label{eq:adsmetr}
ds^2= \frac{d\rho^2}{\rho^2}+\rho^2\eta_{IJ}dx^Idx^J, \qquad
\eta_{IJ} = diag(-1,1,1)\ ,
\end{equation}

We want now to discuss the concept of boundary of our universe.
The first thing to remark is that, strictly speaking, a topological
space has no boundary for its own topology (being at the same time
an open and a closed set).
To provide ${\cal U}$ with a boundary, first of all we have to immerse it
in a ``bigger'' topological space, ${\cal U}'\supset{\cal U}$,
and then we have to determine the boundary $\partial{\cal U}$ with
respect to the surrounding topology.
In this sense, the choice of a boundary is somehow arbitrary,
being arbitrary the choice of $\ {\cal U}'$.

The usual choice of boundary for an Anti--de Sitter space is made in
the following way.
First, we consider $AdS_n$ as a hyperboloid in $\IR ^{n+1}$
(as in section two).
Secondly, we compactify $\IR ^{n+1}$ with a hypersphere $\SS ^n$
given by the limiting points of the straight lines through the origin.
Now we have a hyperboloid immersed in a new topological space,
homeomorphic to the (n+1)--dimensional disk, $B^{n+1}$.
Finally, we call $\partial AdS_n$ the boundary of the hyperboloid
as a subset of this $B^{n+1}$.
Its topology is the same of all the intersections $AdS_n \cap \SS^n(R)$
between the hyperboloid of $\IR^{n+1}$ and the spheres of sufficiently
large radius, $R$: $\SS^1 \times \SS^{n-2}$.

In our case, the actual space--time is the quotient $AdS_4/\ZZ_2$, so
its boundary, even called ``the end of the world'', has the topology
$(\SS^1 \times \SS^2)/\ZZ_2$, where the action of $\ZZ_2$ identifies
points of the form $(\phi,p)$ and $(\phi+\pi,p')$, with $p,p'\in\SS^2$
diametrically opposed points of the sphere.

The action of the isometry group of $\ AdS_4(/\ZZ_2)$, $\ SO(2,3)$, can be
naturally extended to this boundary, which in turn can be provided with
a metric chosen in a set of conformally equivalent ones.
On $(\SS^1 \times \SS^2)(/\ZZ_2),\ SO(2,3)$ acts as the conformal group
for this set of metrics.
In this sense the ``end of the world'' constitutes the support for a
conformal field theory.

Now, in ``physical'' coordinates $\rho,t,w,x$, part of the boundary
is given by the hyperplane $\IR^3(t,w,x)$ at $ \rho\to\infty$.
This subspace inherits from the bulk metric \eqn{eq:adsmetr} a set of
conformally minkowskian metrics, $ds^2=\phi^2(-dt^2+dw^2+dx^2)$ which
can be extended to the conformally invariant compactification of
the minkowskian $\IR^3$.

Finally, we want to remark that the properties of the ``horizon'',
i.e. the site where the membrane lies,  heavily depend on the choice of
compactification.
In ``physical'' coordinates $\rho,t,w,x$, the natural compactification
of the space comprises the set $\rho\to 0$ (i.e. the set of limiting points
of geodesics of the form $(t,w,x)=const,\ \rho\to 0$), which has the
topology of $\IR^3$.
On the other hand, for the topology of the compactification previously
described, all such geodesics converge to the same point of the boundary,
i.e. the membrane shrinks to a single point.

\section*{Appendix C: The supersolvable algebra}
\setcounter{equation}{0}
\renewcommand{\theequation}{C.\arabic{equation}}
Consider the superalgebra $Osp(N \vert 4)$. Its bosonic subalgebra is
$SO(2,3)\times SO(N)$ and it is generated by the momenta $P_a$,
the Lorentz generators $M_{ab}$ and the $SO(N)$ generators $T_{AB}$. 
In this
Appendix $A,B=1,\dots ,N$. The fermionic generators
of the aforementioned algebra are $N$ Majorana spinors $Q^A_\alpha$
in $4$ dimensions.
where $C$ is the charge conjugation matrix in four dimensions
that, in the representation of the Clifford algebra  defined in 
Appendix A,
coincides with $\gamma^0$.

The superspace is defined as the following quotient:
\begin{equation}
AdS^{(N|4)}\,=\,\frac{Osp(N\vert 4)}{SO(1,3)\otimes SO(N)}
\label{supersp}
\end{equation}
and it is spanned by the $4$ coordinates of the $AdS_4$ manifold
and by the $N$ Majorana spinors ($4N$ real components) parametrizing 
the generators
$Q^A_\alpha$ of the superalgebra. It has been shown that the $AdS_4$
manifold admits a solvable description in terms of a $4$ dimensional
solvable Lie algebra $Solv$.
The problem which will be dealt with in the present appendix is
to find
a {\it supersolvable } description of the superspace $AdS^{(N|4)}$, 
that is a decomposition of
$Osp(N\vert 4)$ of the following form:
\begin{eqnarray}
Osp(N\vert 4)\, &=&\, (SO(1,3)\otimes SO(N)\otimes {\cal Q})\oplus 
SSolv,
\label{defissolv}
\end{eqnarray}
where ${\cal Q}$ is a subset of the fermionic generators to 
be defined in
the following. By {\it supersolvable} algebra we mean a superalgebra
for which the $k^{th}$ Lie derivative (defined in terms of the 
supercommutator)
vanishes for a finite $k$.

As pointed out in section $4$, the only price which one has to pay in 
order
to define a supersolvable
algebra $SSolv$ as in eq. (\ref{defissolv}), is to
perform a suitable projection  of the fermionic generators :
\begin{eqnarray}
Q^A_-\,&=&\, {\cal P}_-\cdot Q^A\nonumber \\
Q^A_+\,&=&\, {\cal P}_+\cdot Q^A ,\\
 {\cal P}_\pm^2\, &=&\,{\cal P}_\pm\,\,;\,\,{\cal P}_+\cdot {\cal 
P}_-\,=\,0. \nonumber 
\end{eqnarray}
(differently from the notation used in section $4$, in the present 
appendix
the subscript ``$\pm$'' on fermionic generators denotes the action of 
the the projectors ${\cal P}_\pm$,i.e. generators lying in the upper 
or lower part of the diagram in Figure (1), unless the contrary is 
specified. Therefore 
referring to Figure (1), the fermionic generators in the upper part 
of the diagram are denoted by $Q^A_+$ while the generators $S^A$  in 
the lower part by 
$Q^A_-$. )

Indeed, as already discussed in section $4$, the main idea 
underlying  the 
construction rules of the supersolvable algebra
generating $AdS^{(N|4)}$ as well as the solvable algebra generating 
$AdS$
is that of {\it grading} (figure (1)), i.e. the Cartan 
generator contained in
the coset of $AdS_4$ defines a partition of the isometry generators
into eigenspaces corresponding to positive, negative or null 
eigenvalues
($g_{(\pm 1)},\,sg_{(\pm 1/2)},\,sg_{(0)}$) and the structure of the 
solvable and supersolvable algebras ($Solv$ and $SSolv$) is the 
following:
\begin{eqnarray}
g=SO(2,3) &\rightarrow & g_{(-1)}\oplus g_{(0)}\oplus 
g_{(+1)},\nonumber\\
Solv \, &=&\, \{{\cal C}\}\oplus g_{(-1)},\nonumber\\
\label{grading1}
sg=Osp(N\vert 4)&\rightarrow &g_{(-1)}\oplus sg_{(0)}\oplus 
g_{(+1)}\oplus sg_{(-1/2)}
\oplus sg_{(1/2)},\\
sg_{(0)}\, &=&\,g_{(0)}\oplus SO(N),\nonumber\\
SSolv\, &=&\, \{{\cal C}\}\oplus g_{(-1)}\oplus sg_{(-1/2)}, \nonumber
\end{eqnarray}
where $sg_{(\pm 1/2)} $ represents the grading induced by the Cartan 
generator 
on the fermionic isometries and the eigenspace $sg_{(+ 1/2)} $ not 
entering
the construction of $SSolv$ is the space ${\cal Q}=\{Q^A_+\}$ in eq. 
\eqn{defissolv}
and generates the special conformal transformations. Moreover these 
generators
on the chosen solution of the world volume theory, generate the local
k-supersymmetry transformations ($(1+\bar{\Gamma})/2={\cal P}_+$).

Let us now enter the details of the calculations. As far as the 
$SO(2,3)$
algebra is concerned let us use the following  convention for
 the commutation relations between
its generators $M_{IJ}\,,\, I,J=0,1,2,3, 5$:
\begin{eqnarray}
\left[M^{IJ},M^{KL}\right]\, &=&\, 
-(\eta^{IL}M^{JK}+\eta^{JK}M^{IL}-\eta^{IK}M^{JL}-
\eta^{JL}M^{IK}),\nonumber\\
M^{IJ}\, &=&\,-M^{JI}\,\,,\,\,\eta\,=\,diag\{-,+,+,+,-\}.
\label{so23}
\end{eqnarray}
Let the momenta $P^a$  $(a,b,c=0,\dots, 3)$ be defined as $P^a=M^{a5}$
and the Lorentz generator be $M^{ab}$.

The solvable algebra generating $AdS_4$ has the following form:
\begin{eqnarray}
Solv\, &=&\, \{{\cal C},g_{(-1)}\},\nonumber\\
{\cal C}\, &=&\, M^{25}\,\,;\,\,g_{(-1)}\,=\, \{T,W,X\},\nonumber \\
T\, &=&\, M^{05}-M^{02}\,=\, M^{0-},\\
W\, &=&\, M^{15}-M^{12}\,=\, M^{1-},\nonumber\\
X\, &=&\, M^{35}-M^{32}\,=\, M^{3-},\nonumber
\end{eqnarray}
where the value ``-'' for the index of the $SO(2,3)$ generators refers
to the light--cone notation for the time--like direction $5$ and the
space--like direction $2$.

Using the representation of the Clifford algebra
defined in Appendix A, the spinorial representation of the momenta 
and of the Lorenz generators
$\{P_f^a, M_f^{ab}\}$ consistent with the relations (\ref{so23}) is 
the following :
\begin{eqnarray}
 M_f^{ab}\, &=&\,-\frac{1}{2}\gamma^{ab}\,=\, -\frac{1}{4}\left[
\gamma^a,\gamma^b \right],\nonumber\\
P_f^a\, &=&\,\frac{1}{2}\gamma^5\gamma^a.
\label{spingen}
\end{eqnarray}
With the adopted conventions, the anti--commutator of the 
supersymmetry generators
is given by:
\begin{equation}
\label{superalgebra}
\{Q^A_\alpha,Q^B_\beta\}\, =\, \frac{\delta^{AB}}{2}
(\gamma^{ab}C\gamma^5)_{\alpha\beta} M_{ab}+\delta^{AB}(\gamma^5 
\gamma^a
C\gamma^5)_{\alpha\beta} P_a+{\rm i}(C\gamma^5)_{\alpha
\beta} T^{AB},
\end{equation}
From (\ref{superalgebra})  it is clear that the presence of the 
$SO(N)$ generators $T^{AB}$ on the right hand side of the 
anticommutator 
between fermionic generators
is an obstacle for the definition of a solvable superalgebra.
As it will be shown that term disappears once
the projection on $sg_{(-1/2)}$ is performed on the fermionic 
generators.

Now let us define the grading for the fermionic generators with 
respect to ${\cal C}$. Since the
adjoint action of ${\cal C}$ on $Q^a$ is represented by
 the matrix ${\cal C}^f=P_f^2=\gamma^5 \gamma^2/2$ acting on the 
fermions,
the projector on the spaces $sg_{(\pm 1/2)}$ is given by:
\begin{eqnarray}
{\cal P}_\pm \, &=&\, \frac{1}{2}(\unity \pm 
\gamma^5\gamma^2),\nonumber\\
sg_{(\pm 1/2)} \, &=&\, \{Q^A_{\pm}\}\, =\, \{{\cal P}_\pm Q^A\}.
\label{projec}
\end{eqnarray}
It is straightforward to verify that such a projection is compatible 
with the Majorana condition.
The solvable superalgebra has then the following content:
\begin{equation}
SSolv\,=\, Solv\oplus sg_{(-1/2)}\,=\,\{{\cal C}, T, W,X\}\oplus 
\{Q^A_-\}.
\end{equation}
The fact that it closes follows form the rules (\ref{superalgebra}) 
and from the grading of the
generators: the anticommutator of two $Q^A_-$ has charge
$-1$ with respect to ${\cal C}$ and therefore is expressed only in 
terms
of the $T, W,X$ generators and so on.
Let us define the fermionic eigenmatrices
$T^f_\pm, W^f_\pm, X^f_\pm$ of ${\cal C}^f=\frac{1}{2}\gamma^5 
\gamma^2=P_f^2$:
\begin{eqnarray}
\left[{\cal C}^f,T^f_\pm\right]\, =\,\pm T^f_\pm,  &
\left[{\cal C}^f,W^f_\pm\right]\, =\,\pm W^f_\pm,  &
\left[{\cal C}^f,X^f_\pm\right]\, =\,\pm X^f_\pm,\nonumber\\
 T^f_\pm\, =\,P_f^0\pm M_f^{02}, &
W^f_\pm\, =\,P_f^1\pm M_f^{12},  &
X^f_\pm\, =\,P_f^3\pm M_f^{32}.
\label{TWXf}
\end{eqnarray}
After some gamma-algebra one finds for the anticommutator of two $Q_-$
the following expression:
\begin{equation}
\{Q_-,Q_-\}\, =\, -T^f_- C\gamma^5 T+ W^f_- C\gamma^5 W+ X^f_- C\gamma^5 X,
\end{equation}
where we used the property: ${\cal P}_-\{T^f_-,W^f_-,X^f_-\}  C{\cal 
P}_-=
\{T^f_-,W^f_-,X^f_-\}  C$. 

As far as the anti--commutation relations between the fermionic 
generators 
and the
$AdS$ isometry generators are concerned, taking into account 
 \eqn{TWXf} they may be rewritten in the following way,
\begin{equation}
\label{TWXQ}
\left[T,Q^A\right]\, =\,T^f_+ Q^A, \qquad
\left[W,Q^A\right]\, =\,W^f_+ Q^A, \qquad
\left[X,Q^A\right]\, =\,X^f_+ Q^A.
\end{equation}
Projecting eqs. (\ref{TWXQ}) on $sg_{(-1/2)}$, using the property 
${\cal P}_- \cdot T^f_+={\cal P}_- \cdot W^f_+={\cal P}_- \cdot 
X^f_+=0$
(since the generators $\{  T^f_+,W^f_+,X^f_+\}$ shift the eigenvalue
of ${\cal C}^f$ by $+1$) one obtains:
\begin{equation}
\label{T}
\left[T,Q^A_-\right]\, =\,0,\qquad
\left[W,Q^A_-\right]\, =\,0,\qquad
\left[X,Q^A_-\right]\, =\,0.
\end{equation}
Defining for simplicity 
$Z^\mu=\{T,W,X\}\,;\,Z_f^\mu=\{T^f_-,W^f_-,X^f_-\}$,
it is now possible to
write the algebraic structure of $SSolv$:
\begin{eqnarray}
\left[{\cal C}, Z_\mu \right]\, &=&\, -Z_\mu,\nonumber\\
\left[Z_\mu, Z_\nu \right]\, &=&\,0,\nonumber\\
\label{SSolv}
\left[Z_\mu, Q^A_-\right]\, &=&\,0,\\
\left[{\cal C}, Q^A_- \right]\, &=&\,-\frac{Q^A_-}{2},\nonumber\\
\{Q^A_-, Q^B_-\}\, &=&\,\delta^{AB}h^{\mu \nu}(Z^f_{\mu}C\gamma^5) 
Z_\nu \,\,\,\,\,h=diag\{-,+,+\}. \nonumber
\end{eqnarray}
Using \eqn{matr1}--\eqn{matr5} conventions, the spinorial 
representation 
${\cal C}^f$ of the Cartan generator has the form:
\begin{eqnarray}
{\cal C}^f\, &=&\,\frac{1}{2} \left(\begin{tabular}{cc} $-\unity$ & 
$0$ \\ 
$0$ & $\unity$\end{tabular} \right)
\end{eqnarray}
and therefore the projectors are:
\begin{eqnarray}
{\cal P}_+\, &=&\,\left( \matrix{ 0 & 0  \cr 0 & \unity 
}\right),\,\,\,\,{\cal P}_-\,=\,\left( \matrix{ \unity & 0  \cr 0 & 0 
}\right).
\end{eqnarray}
In the spinorial representation of $Solv$, besides $C^f$ there are 
the {\it maximal abelian} generators $T^f_-,W^f_-,X^f_-$ appearing 
in  the coset representative through the following combinations:
\begin{eqnarray}
{\cal \sigma}_{\perp} &=& \frac{1}{\sqrt{2}}X^f_-=\frac{1}{2} 
\left(\begin{tabular}{cc} $0$ & $\sqrt{2}\s^3$ \\ $0$ & $0$
\end{tabular} \right), \nonumber\\
{\cal \sigma}_{+} &=&  \frac{1}{2}(-T^f_-+W^f_-)=\frac{1}{2} 
\left(\begin{tabular}{cc} $0$ & $- \unity + \s^1$ \\ $0$ & $0$
\end{tabular} \right),\nonumber\\
{\cal \sigma}_{-} &=& \frac{1}{2}(-T^f_--W^f_-)=\frac{1}{2} 
\left(\begin{tabular}{cc} $0$ & $- \unity - \s^1$ \\ $0$ & $0$
\end{tabular} \right).
\end{eqnarray}
where the ``$\pm$'' label on the operators ${\cal \sigma}$ refers to
the light--cone index related to the coordinates $t$ and $w$.
 
Promoting all the bosonic matrices to graded ones preserving 
\eqn{orfan27},
the bosonic factor of the group representative is
\begin{equation}
L_B = Exp(\sqrt{2}x{\cal \sigma}_{\perp}+
t({\cal \sigma}_{+}+{\cal \sigma}_{-})+
w({\cal \sigma}_{+}-{\cal \sigma}_{-}))\,e^{a \, {\cal C}^f},
\end{equation}
and the coset representative of the superspace in the supersolvable 
parametrization has the form $L = L_F L_B$ where
\begin{equation}
L_{F} = exp\left( \theta_1^A Q_1^A + \theta_2^A  Q_2^A \right).
\end{equation}
The left--invariant one--form is therefore given by
\begin{equation}
\label{liform}
\Omega = L^{-1}dL=\Omega_B + L_B^{-1} \Omega_F L_B.
\end{equation}
From the left invariant form (\ref{liform}), we can finally
obtain the vielbeins through the following projections:
\begin{eqnarray}
E^0 &=& \frac{1}{2}{\rm Tr}\left(\gamma^5\gamma^0 
\Omega\right),\nonumber\\
E^i &=& \frac{1}{2}{\rm Tr}\left(\gamma^5\gamma^i \Omega\right).
\end{eqnarray}

We have then parametrized our space with \eqn{solvmet} metric.
To go back to the horospherical coordinates \eqn{nearhor}, 
we have first to rescale
all the vielbeins by a $R/2$ factor and then to reabsorb this factor 
in the 
$\rho$ definition.
The final parametrization is then given by \eqn{param1}--\eqn{param5} 
reported in section \eqn{quattro}.

\end{document}